\documentclass[usenatbib,letterpaper,hyperref]{mn2e}
\usepackage[labelfont=bf,labelsep=period]{caption}
\usepackage[totalwidth=480pt, totalheight=680pt]{geometry}
\usepackage{hyperref}
\usepackage{times}
\usepackage{epsfig}
\usepackage{amsmath}
\usepackage{amssymb}
\usepackage{url}
\usepackage{aas_macros}
\def\eqsim{\mathrel{\raise0.35ex\hbox{$\scriptstyle =$}\kern-0.6em
    \lower0.40ex\hbox{{$\scriptstyle \sim$}}}}
\def\gtrsim{\mathrel{\raise0.35ex\hbox{$\scriptstyle >$}\kern-0.6em
    \lower0.40ex\hbox{{$\scriptstyle \sim$}}}}
\def\lesssim{\mathrel{\raise0.35ex\hbox{$\scriptstyle <$}\kern-0.6em
    \lower0.40ex\hbox{{$\scriptstyle \sim$}}}}
\def\HI{H\,{\sc i}}
\def\AppPPSMaps{A}
\def\AppExtRes{B}
\def\AppAltGas{C}
\def\AppExtSim{D}
\def\AppTables{E}
\def\AppCorners{F}

\defcitealias{2013MNRAS.432..336W}{W13}
\defcitealias{2016MNRAS.463.3083O}{OH16}

\title[Stripping and quenching of satellites]{A homogeneous measurement of the delay between the onsets of gas stripping and star formation quenching in satellite galaxies of groups and clusters}
\author[K. A. Oman et al.]{
  \newauthor Kyle A. Oman$^{1,2}$\thanks{kyle.a.oman@durham.ac.uk}, Yannick M. Bah\'e$^{3}$, Julia Healy$^{2,4}$, Kelley M. Hess$^{2,5}$, \newauthor Michael J. Hudson$^{6,7,8}$, Marc A. W. Verheijen$^{2}$\\
  $^{1}$ Institute for Computational Cosmology, Department of Physics, Durham University, South Road, Durham DH1 3LE, United Kingdom\\
  $^{2}$ Kapteyn Astronomical Institute, University of Groningen, Postbus 800, NL-9700 AV Groningen, The Netherlands\\
  $^{3}$ Leiden Observatory, Leiden University, PO Box 9513, NL-2300 RA Leiden, the Netherlands\\
  $^{4}$ Department of Astronomy, University of Cape Town, Private Bag X3, Rondebosch 7701, South Africa\\
  $^{5}$ ASTRON, the Netherlands Institute for Radio Astronomy, Postbus 2, NL-7990 AA Dwingeloo, the Netherlands\\
  $^{6}$ Department of Physics and Astronomy, University of Waterloo, Waterloo, ON N2L 3G1, Canada\\
  $^{7}$ Waterloo Centre for Astrophysics, University of Waterloo, Waterloo, ON N2L 3G1, Canada\\
  $^{8}$ Perimeter Institute for Theoretical Physics, Waterloo, ON N2L 2Y5, Canada\\
}
\date{\today}

\begin{document}
\label{firstpage}
\maketitle

\begin{abstract} 
  We combine orbital information from N-body simulations with an analytic model for star formation quenching and SDSS observations to infer the differential effect of the group/cluster environment on star formation in satellite galaxies. We also consider a model for gas stripping, using the same input supplemented with \HI\ fluxes from the ALFALFA survey. The models are motivated by and tested on the Hydrangea cosmological hydrodynamical simulation suite. We recover the characteristic times when satellite galaxies are stripped and quenched. Stripping in massive ($M_{\rm vir}\sim 10^{14.5}\,{\rm M}_\odot$) clusters typically occurs at or just before the first pericentric passage. Lower mass ($\sim10^{13.5}\,{\rm M}_\odot$) groups strip their satellites on a significantly longer (by $\sim3\,{\rm Gyr}$) timescale. Quenching occurs later: Balmer emission lines typically fade $\sim3.5\,{\rm Gyr}$ ($5.5\,{\rm Gyr}$) after first pericentre in clusters (groups), followed a few hundred ${\rm Myr}$ later by reddenning in $(g-r)$ colour. These `delay timescales' are remarkably constant across the entire satellite stellar mass range probed ($\sim10^{9.5}$--$10^{11}\,{\rm M}_\odot$), a feature closely tied to our treatment of `group pre-processing'. The lowest mass groups in our sample ($\sim10^{12.5}\,{\rm M}_\odot$) strip and quench their satellites extremely inefficiently: typical timescales may approach the age of the Universe. Our measurements are qualitatively consistent with the `delayed-then-rapid' quenching scenario advocated for by several other studies, but we find significantly longer delay times. Our combination of a homogeneous analysis and input catalogues yields new insight into the sequence of events leading to quenching across wide intervals in host and satellite mass.
\end{abstract}
\begin{keywords}
galaxies: clusters: general -- galaxies: groups: general -- galaxies: evolution
\end{keywords}

\section{Introduction}
\label{SecIntro}

Galaxies do not exist in isolation. They occupy the various environments embedded within the cosmic web -- nodes, filaments, walls and voids, in rough order of decreasing galaxy density -- and may be either a locally dominant `central' object, or a `satellite' embedded in a group or cluster. Many of the processes which shape the build-up to a present-day galaxy are sensitive to its surroundings. Gas accretion, for instance, is suppressed in higher density environments where galaxies typically move faster relative to a hotter ambient medium. Satellite galaxies may lose material to the ram pressure felt as they plough through the coronal gas of their host systems, or to tides. Mergers become more common with increasing density, but drop off again once the typical relative velocities increase enough to make fly-by encounters more common. Some environmental processes couple non-linearly. For instance, `harassing' fly-by encounters make the galaxies involved more susceptible to stripping mechanisms; ram pressure compresses gas as well as stripping it, which may trigger increased star formation, in turn feeding energy back into the ISM and driving outflows, further accelerating the loss of gas \citep[see also][sec.~4 for a more detailed overview of the outline above]{2006PASP..118..517B}.

\subsection{Quenching and the environment}

Despite the numerous physical processes involved and complex interactions between them, some simple broad trends emerge: galaxy populations in denser environments include proportionally more early-types \citep{1980ApJ...236..351D}, and fewer star-forming galaxies \citep{2004ApJ...615L.101B,2004ApJ...601L..29H}. Not all galaxies are equally sensitive to environment: in a seminal study, \citet{2010ApJ...721..193P} showed that the dependence of star formation (as traced by colour) on density is at least approximately separable from its dependence on stellar mass. Furthermore, star formation in lower mass galaxies is only shut down -- `quenched' -- in the highest density environments, while the most massive galaxies ($M_\star \gtrsim 10^{11}\,{\rm M}_\odot$) tend to be non-star-forming (`passive') independent of local density.

Given the large number of potentially important physical processes at play, it is interesting to ask which are dominant in shaping the evolution of galaxies at different stages of their assembly and in environments of different densities. One possible approach is to look for relatively sharp transitions in the properties of the galaxy population. For instance, a transition in the typical colour of satellite galaxies as a function radial separation from a host object can be statistically related to the time spent in a high density environment, allowing an inference on the timing and duration of the transition along the satellite orbits. These timescales can then be compared to theoretical predictions for the timescales on which the various processes should operate, linking each observed transition to a physical explanation. This general approach has been used by many authors, for a few notable examples see e.g. \citet{2000ApJ...540..113B,2004MNRAS.353..713K,2010ApJ...721..193P,2013MNRAS.432..336W,2015ApJ...806..101H,2017ApJ...835..153F}; see also other specific examples cited below.

In this work we will focus our attention specifically on the regions of highest galaxy density -- galaxy groups and clusters. It has been shown that the influence of the group/cluster environment on star formation extends out to several ($\sim 5$) virial\footnote{Throughout this work, we define `virial' quantities at an overdensity corresponding to the solution for a spherical top-hat perturbation which has just virialized, see e.g. \citet{1998ApJ...495...80B}. The virial overdensity at $z=0$ is $\sim 360\times$ the background density $\Omega_{\rm m}\rho_{\rm c}$, where $\rho_{\rm c}$ is the critical density for closure and $\Omega_{\rm m}$ is the cosmic matter density in units of the critical density. For readers accustomed to a virial overdensity of $200\times$ the critical density, approximate conversions are $M_{200}/M_{\rm vir}\approx0.81$ and $r_{200}/r_{\rm vir}\approx0.73$, assuming a typical concentration parameter for cluster-scale systems.} radii \citep{1999ApJ...527...54B,2010MNRAS.404.1231V,2012MNRAS.420..126L,2012ApJ...757..122R}. On their first orbit through a cluster, some galaxies may reach apocentric distances of up to $\sim 2.5\,r_{\rm vir}$ \citep[e.g.][]{2004A&A...414..445M,2013MNRAS.431.2307O}, and such `backsplash' objects are required to explain the suppression of star formation in galaxies in the immediate vicinities of groups/clusters \citep{2014MNRAS.439.2687W}.

\subsection{Group pre-processing}

That star formation is reduced relative to more isolated galaxies to much larger radii indicates that star formation is sensitive to more than the direct influence of the cluster. For instance, galaxies may feel ram pressure as they fall through the filaments feeding material onto the cluster \citep{2013MNRAS.430.3017B}. The hierarchical clustering of galaxies also implies that groups are proportionally more common around clusters. Star formation in many cluster satellites is thus likely affected by processes particular to the group environment long before they actually reach the cluster itself \citep{2018ApJ...866...78H}, especially within filaments \citep{2020A&A...635A.195G}. Similarly, group satellites fall into their hosts as members of smaller groups. This broad notion has become known as `group pre-processing', and is recognized as a crucial ingredient in models seeking to explain the environmental dependence of star formation \citep[e.g.][]{2004PASJ...56...29F,2014MNRAS.440.1934T,2014MNRAS.442..406H,2015MNRAS.448.1715J,2015ApJ...806..101H,2019MNRAS.488..847P,2020A&A...635A.195G,2020ApJS..247...45R,2020MNRAS.tmp.2921D}.

\subsection{Projected phase space}

The three observationally accessible `projected phase space' (PPS) coordinates of a satellite, i.e. its on-sky position and line-of-sight velocity offsets from its host system, correlate with parameters describing its orbit, such as the time since infall, or distance of closest approach \citep{2005MNRAS.356.1327G,2011MNRAS.416.2882M,2013MNRAS.431.2307O,2019MNRAS.484.1702P}. Note that the sign of the line-of-sight velocity usually carries no information: in the absence of high-precision distance measurements, it is impossible to discriminate between e.g. a foreground satellite receding toward a background host (negative radial velocity with respect to the host), or a background satellite receding away from a foreground host (positive radial velocity with respect to the host). The possible orbits corresponding to a given set of PPS coordinates can be inferred by sampling in simulations the orbits of satellites with similar PPS coordinates, suitably normalized, around broadly similar hosts; the dependence of the resulting orbital distribution on the host mass and the ratio of the satellite and host masses are relatively weak \citep{2013MNRAS.431.2307O}.

Analyses of the PPS coordinates of galaxies and their correlations with other galaxy properties have led to many inferences on the environmental influence of groups and clusters, for instance: that star formation is nearly completely quenched after a single passage through a rich cluster \citep{2011MNRAS.416.2882M}; that the predicted strength of the ram-pressure force as a function of PPS coordinates is anti-correlated with the PPS distribution of star-forming galaxies \citep[][see also \citealp{2019MNRAS.484.3968A,2020MNRAS.495..554R}]{2014MNRAS.438.2186H}; that atomic hydrogen-deficient blue galaxies are on average further along their orbits within clusters than atomic hydrogen-rich blue galaxies, suggesting that they have been ram-pressure stripped \citep{2015MNRAS.448.1715J}; that warm dust is also ram-pressure stripped \citep{2016ApJ...816...48N}; that galaxies are morphologically segregated in PPS in the Coma cluster, with late-types preferentially exhibiting tails of stripped H~$\alpha$-emitting gas \citep{2018A&A...618A.130G}; that galaxies are mass-segregated in clusters \citep{2020arXiv201005304K}; that galaxies exhibiting the most prominent tails in H~$\alpha$ are preferentially found at low projected position offset and high projected velocity offset, consistent with being at their first pericentre \citep{2018MNRAS.476.4753J}; that the ages of stellar populations are strongly correlated with PPS coordinates, and that morphological transformation lags the shutdown in star formation \citep{2019MNRAS.486..868K}; that the shapes of SED-fitting based galaxy star formation histories correlate with PPS coordinates \citep{2019ApJ...876..145S}.

\subsection{Quenching timescales}

In this work we focus particularly on the timescales associated with the quenching of satellites in groups and clusters. There is a growing consensus in the literature that quenching of satellites of $M_\star\gtrsim 10^9\,{\rm M}_\odot$ in low-redshift groups ($M_{\rm vir}\gtrsim 10^{13}\,{\rm M}_\odot$) proceeds in a `delayed-then-rapid' fashion, with star forming galaxies continuing their activity for several Gyr after infall into a host, followed by an abrupt cessation on a much shorter timescale. This was initially suggested based on semi-analytic models \citep{2012MNRAS.423.1277D} and measured shortly thereafter \citep[][hereafter \citetalias{2013MNRAS.432..336W}]{2013MNRAS.432..336W}, followed by corroboration by several other authors (e.g. \citealp{2014MNRAS.440.1934T,2015ApJ...806..101H}; \citealp{2016MNRAS.463.3083O}, hereafer \citetalias{2016MNRAS.463.3083O}; \citealp{2019A&A...621A.131M,2020ApJS..247...45R,2020arXiv200811663A}). Whether different measurements agree in detail, for instance regarding the delay timescale, is more difficult to assess. Each study makes a different set of modelling assumptions, and in most cases chooses a different reference `infall time', e.g. infall is defined at different radii, or as first infall into any host system rather than infall into the present host system. Given the diversity in satellite orbits, translating between different definitions is not straightforward. In this study we use the time of the first pericentric passage, rather than an `infall time', as our primary reference time. This is the time where both the tidal and ram-pressure forces acting on the satellite are expected to first peak and so this definition may simplify the interpretation of our measurement, though we note that `infall', however defined, is likely also relevant in that this is approximately the time when accretion of fresh gas onto the satellite would be expected to cease.

The physical interpretation of the `delayed-then-rapid' measurement is debated. While it is generally agreed that accretion of fresh gas should cease around the time a satellite enters the intra-group/cluster medium of its host, the extent to which the remaining gas supply is depleted by continued star formation (and associated feedback-driven winds) versus removed by ram pressure is unclear. While some authors argue that a starvation\footnote{Some authors distinguish between `starvation', the consumption of gas in the absence of accretion, and `strangulation', the stripping of hot gas truncating cooling into colder phases. We do not attempt to make this distinction, and use `starvation' to encompass both.} model alone can adequately explain the measurements \citep[e.g. \citetalias{2013MNRAS.432..336W};][]{2014MNRAS.440.1934T}, others contend that ram-pressure stripping (RPS) plays a significant role \citep{1999ApJ...516..619F,2013ApJ...775..126H,2015MNRAS.448.1715J,2019ApJ...873...42R,2020ApJS..247...45R}. Curiously, some analyses of hydrodynamical simulations strongly favour a RPS dominated scenario \citep[][see also Sec.~\ref{SubsecModelMotivation} below]{2015MNRAS.447..969B,2019MNRAS.488.5370L}, though others argue that starvation alone may be sufficient, particularly for low-mass galaxies \citep{2017MNRAS.466.3460V}. \citet[][see also \citealp{2016MNRAS.456.1115B}]{2017MNRAS.470.4186B} also caution that, due to their limits in terms of length resolution (as limited by the gravitaional softening) and/or temperature (where a cooling floor is imposed in the ISM model), current hydrodynamical simulations of clusters likely overpredict the efficiency of RPS.

Further insight into the physics of quenching can be gleaned by considering the redshift and host mass dependences of the delay timescale, in particular. \citet{2013MNRAS.431.1090M,2014MNRAS.438.3070M} and \citet{2014ApJ...796...65M} find that a `delayed-then-rapid' scenario also seems to hold for galaxy groups and clusters at $z\sim 1$, but that the delay timescale must be much shorter \citep[but see also][who infer a somewhat longer timescale]{2017ApJ...835..153F}. This argues either for an increased importance of RPS \citep{2014ApJ...796...65M,2015MNRAS.447..969B}, or strong, wind-driven outflows \citep[][see also \citealp{2019MNRAS.490.1231L} who argue against the importance of RPS]{2014MNRAS.442L.105M,2016MNRAS.456.4364B}. While for group and cluster satellites of $M_\star\gtrsim10^9\,{\rm M}_\odot$ the delay timescale decreases with increasing stellar mass, \citet[][see also \citealp{2019arXiv190604180F,2020arXiv200307006M}]{2015MNRAS.454.2039F} point out that this trend must eventually turn over at lower host and/or satellite masses, as the satellites of the Milky~Way seem to have very short delays between infall and quenching. They interpret this as evidence of a transition from a starvation-dominated scenario at higher masses to a RPS dominated scenario for the Milky~Way satellites. The interpretation of a measurement of a very long ($\sim 10\,{\rm Gyr}$) delay time for dwarfs in group-mass hosts \citep{2014MNRAS.442.1396W} remains an open question (but see Sec.~\ref{SubsecInterpret} below for an interpretation of our qualitatively similar result).

\subsection{Outline}

In this work we aim to measure timescales pertinent to quenching in groups and clusters. We use a homogeneous modelling process and, as far as possible, homogenous input data drawn from the Sloan Digital Sky Survey (SDSS), in order to enable a straightforward comparison across $\sim 1.5$ decades in satellite stellar mass and $\sim 3$ decades in host (total) mass. Our model is broadly motivated by the `delayed-then-rapid' paradigm, but allows for substantial scatter in the timing of the `rapid' phase for individual galaxies within a population. The form of our model is further inspired by, but not explicitly tied to, an analysis of quenching in the Hydrangea cluster zoom-in cosmological hydrodynamical simulation suite \citep[][see also \citealp{2017MNRAS.471.1088B}]{2017MNRAS.470.4186B}. We adopt a similar methodology to \citetalias{2016MNRAS.463.3083O}, using a large cosmological N-body simulation to infer the probable orbits of observed satellites based on their PPS coordinates. A crucial difference with that study, however, is that we build our model around the probability distribution for the time of the first pericentric passage, rather than the time of first infall into the final host system.

While many studies explicitly model group pre-processing, in this work we adopt a qualitatively different approach, following \citetalias{2016MNRAS.463.3083O}. Rather than use a galaxy population well-removed from the group or cluster under consideration as a reference (often termed `field') sample, we compare group/cluster members to galaxies in the immediate vicinity of the host system. We thus aim to isolate and measure the differential effect of the host on the star formation of galaxies relative to what it would have been had they not fallen into the host but otherwise kept the same evolutionary path up to that point.

We also aim to measure the timescale for the depletion of H\,{\sc i} gas (either through conversion into stars, stripping, or a change in phase), using as far as possible the same methodology and a subsample of the same input galaxies where the SDSS overlaps with the Arecibo Legacy Fast ALFA survey (ALFALFA). The combination of constraints on the gas-stripping and star formation-quenching timescales constitutes a powerful probe of the physics regulating star formation in satellite galaxies. For instance, an abrupt decline in H\,{\sc i} content synchronized with a decline in SFR would argue strongly in favour of rapidly-acting RPS, especially if occurring near pericentre, whereas a more gradual decline in H\,{\sc i} content accompanied by sustained star formation would be more consistent with a starvation scenario.

This paper is organised as follows. In Sec.~\ref{SecData} we outline the survey catalogues we use as input. The SDSS data used for our quenching analysis are described in Sec.~\ref{SubsecDataQuenching}; the additional ALFALFA data used to supplement the SDSS catalogue for our stripping analysis is described in Sec.~\ref{SubsecDataStripping}. In Sec.~\ref{SecSims} we outline the simulation datasets we use to motivate the form of our model for quenching (Sec.~\ref{SubsecHydrangea}) and infer the orbital parameters of obseved galaxies given their PPS coordinates (Sec.~\ref{SubsecNbody}). We describe our quenching model in Sec.~\ref{SecModel}, including its motivation (Sec.~\ref{SubsecModelMotivation}), formal definition and statistical analysis methodology (Sec.~\ref{SubsecDefinitionFitting}), and tests of our ability to accurately recover model parameter values (Sec.~\ref{SubsecModelTests}). We present our measurements of the quenching and stripping timescales as a function of stellar mass and host mass in Sec.~\ref{SecResults}, discuss our interpretation thereof in Sec.~\ref{SecDiscussion}, and summarize in Sec.~\ref{SecConc}.

\section{Observed galaxy sample}
\label{SecData}

\subsection{Sample for quenching analysis}
\label{SubsecDataQuenching}

We make use of the SDSS Data Release~7 \citep{2009ApJS..182..543A} catalogue, supplemented with star formation rates \citep{2004MNRAS.351.1151B,2007ApJS..173..267S} and stellar masses \citep{2014ApJS..210....3M}. We use the spectroscopic sample of galaxies, which introduces incompleteness in the sample at the 10~per~cent level globally (likely higher in dense clusters) due to fibre collisions. At fixed galaxy density, this bias is not strongly dependent on e.g. colour, such that the effect on our statistical analysis should be minimal (see Sec.~\ref{SubsubsecCompleteness} for a more detailed discussion of which types of biases are likely to affect our analysis). We discard all galaxies with $m_r > 17.5$, which yields a complete (except for fibre collisions) magnitude limited sample \citep{2002AJ....124.1810S}\footnote{See also \url{https://classic.sdss.org/dr7/products/general/target_quality.html}.}. Within a given group, all galaxies are at approximately the same distance, so a magnitude limit translates approximately to a stellar mass limit \citep[the $r$-band is a reasonable tracer of total stellar mass;][]{1999ASPC..163...28M}. Since we fit our model (Sec.~\ref{SubsecDefinitionFitting}) to data in narrow bins in stellar mass, the net effect of the magnitude limit is simply to change the number of groups contributing to any given stellar mass bin, with more distant groups dropping out of the sample for bins at lower stellar masses. Thus, each fit is actually performed on an approximately \emph{volume}-limited sample of galaxies, with the volume covered varying with stellar mass.

We further prune our sample of galaxies, removing those with $M_\star < 10^{9.5}\,{\rm M}_\odot$. This is because these low-mass galaxies have a relative $r$-band magnitude-dependent bias in their colour distribution, such that there are relatively more faint red galaxies in the catalogue, which could significantly bias our analysis. This issue is discussed further in Sec.~\ref{SubsubsecCompleteness}.

In order to obtain a sample with a wide range in host halo mass, we select satellite galaxy candidates around groups/clusters from the \citet{2007MNRAS.379..867V} and \citet{2017MNRAS.470.2982L} group catalogues: the group virial masses of these two catalogues peak at $\sim 3\times10^{14}$ and $3\times10^{13}\,{\rm M}_\odot$, respectively. We discard the small number of groups with redshifts $z < 0.01$, which typically have very bright members not covered by SDSS spectroscopy. Although the algorithms used to construct the two group catalogues are quite different -- \citet{2007MNRAS.379..867V} search for overdensities of galaxies which share similar colours, while \citet{2017MNRAS.470.2982L} use a friends-of-friends-based approach -- our methodology is minimally sensitive to the any resulting differences as we use only the group centres, redshifts, and halo masses from these catalogues, and not galaxy membership information (the reasons for this are further elaborated below). We derive group velocity dispersions (Eq.~\ref{EqVdisp}, below) and use these to normalize the velocity offsets of satellites from their hosts. The velocity dispersion which we calculate is closest to the dark matter particle velocity dispersion of the system \citep[within about 10~per~cent, see][esp. their table~1]{2013MNRAS.430.2638M}, which is the velocity dispersion which we use to normalize the velocity offsets of satellite haloes in our N-body simulations (see Sec.~\ref{SubsecNbody}), making these two sets of normalized coordinates mutually compatible.

We calculate halo masses for the host systems following \citet[][eq.~1]{2007MNRAS.379..867V} and \citet[][eq.~4]{2017MNRAS.470.2982L}, and convert these to virial masses by assuming a \citet{1996ApJ...462..563N} density profile and the mean mass-concentration relation of \citet{2014MNRAS.441..378L}, accounting throughout for differences in the assumed cosmologies. The virial radii follow from the mean enclosed density as $r_{\rm vir}=\left(\frac{3}{4\pi}\frac{M_{\rm vir}}{\Delta_{\rm vir}(z)\Omega_{\rm m}(z)\rho_{\rm crit}(z)}\right)^{\frac{1}{3}}$, where $\Delta_{\rm vir}$ is the virial overdensity in units of the mean matter density $\Omega_{\rm m}\rho_{\rm crit}$ and $\rho_{\rm crit}=\frac{3H^2}{8\pi G}$. Finally, we estimate the velocity dispersions of the groups following \citet[][but accounting for the redshift dependence, see \citealp{1998ApJ...495...80B}]{2006A&A...456...23B} as:
\begin{equation}
  \frac{\sigma_{1{\rm D}}}{{\rm km}\,{\rm s}^{-1}} = \frac{0.0165}{\sqrt{3}}\left(\frac{M_{\rm vir}}{{\rm M}_\odot}\right)^{\frac{1}{3}}\left(\frac{\Delta_{\rm vir}(z)}{\Delta_{\rm vir}(0)}\right)^{\frac{1}{6}}(1+z)^{\frac{1}{2}}\label{EqVdisp}
\end{equation}

\begin{figure*}
  \includegraphics[width=\textwidth]{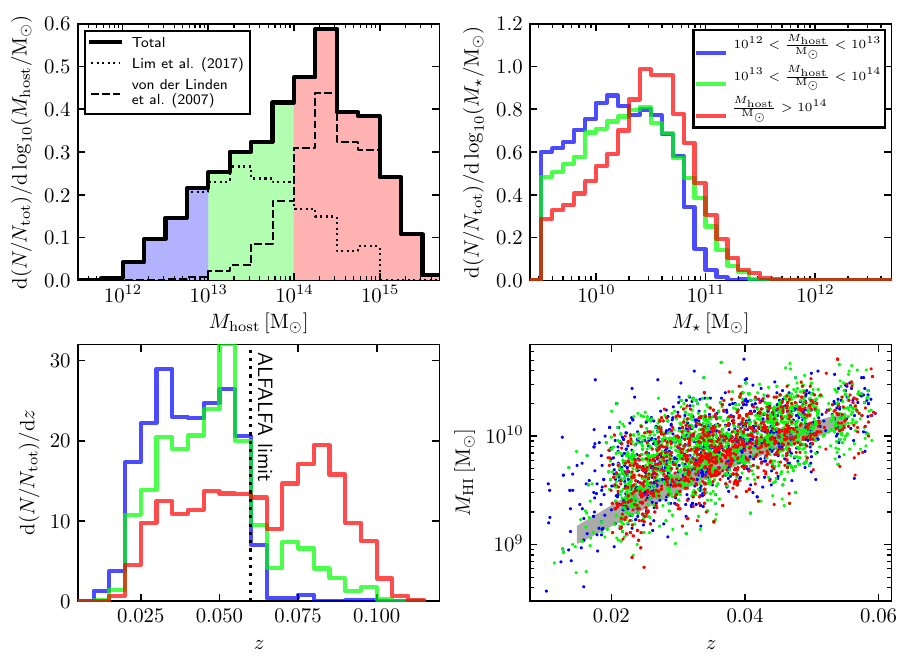}
  \caption{Overview of the observational samples. \emph{Upper left}: Normalized host virial mass distribution for the combined matches to the \citet{2007MNRAS.379..867V} and \citet{2017MNRAS.470.2982L} group catalogues (solid line). The matches to the individual catalogues are shown with broken lines, as labelled. The histogram reflects the number of satellite candidates around hosts of each mass, not the number of host systems. The low-, intermediate- and high-mass host sample ranges are highlighted in blue, green and red, respectively. \emph{Upper right}: Normalized satellite stellar mass distribution in each host mass bin, as labelled. We truncate the sample at $M<10^{9.5}\,{\rm M}_\odot$ (see Sec.~\ref{SubsecDataQuenching}). \emph{Lower left}: Normalized redshift distribution, weighted by satellite candidate count, of hosts for each host mass bin. The redshift limit of $z\sim0.06$ of the \citep{2018ApJ...861...49H} source catalogue is marked with the vertical dashed line; for brevity, we do not show the distributions for the ALFALFA cross-matched galaxy sample. \emph{Lower right}: Redshifts and \HI\ masses of ALFALFA-detected satellite candidates. The gray band shows the interquartile range of upper limit estimates for non-detections (see Sec.~\ref{SubsecDataStripping}).}
  \label{FigObsOverview}
\end{figure*}

The host halo mass, satellite stellar mass and redshift distributions, and the \HI\ masses vs. redshift for ALFALFA-detected satellite candidates, for the two samples are shown in Fig.~\ref{FigObsOverview}. The host halo mass and redshift distributions are weighted by the number of candidate members used in our analysis (see below): they represent relative numbers of galaxies, not of groups.

\begin{figure*}
  \includegraphics[width=\textwidth]{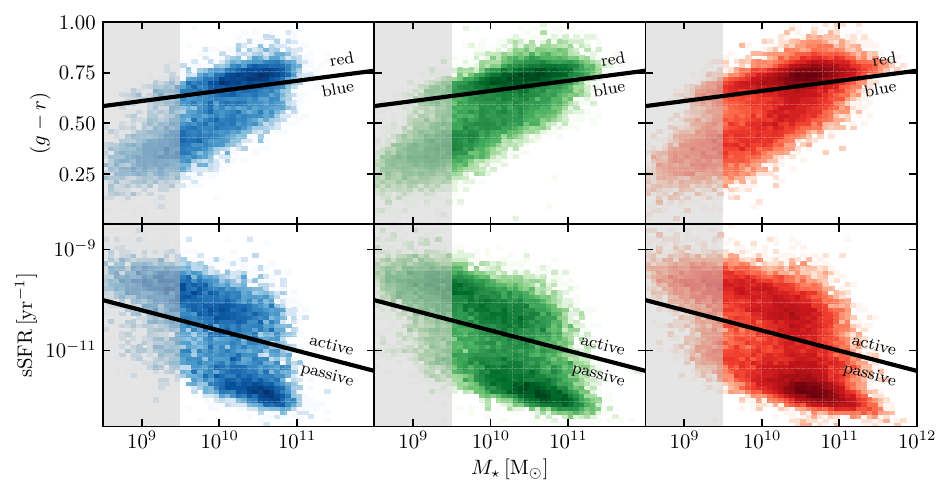}
  \caption{Distribution of galaxies in the low- (left panels), intermediate- (centre panel) and high-mass host samples in the $(g-r)$ colour--stellar mass (upper panels) and sSFR--stellar mass (lower panels) planes. The pixel colour is logarithmically scaled. The solid lines show our adopted divisions between the red and blue populations, $(g-r)=0.05\log_{10}(M_\star/{\rm M}_\odot)+0.16$, and the active and passive populations, ${\rm sSFR}/{\rm yr}^{-1}=-0.4\log_{10}(M_\star/{\rm M}_\odot)-6.6$. Galaxies with $M_\star<10^{9.5}\,{\rm M}_\odot$ are excluded from our analysis (see Sec.~\ref{SubsecDataQuenching}) -- this region is shaded in gray.}
  \label{FigCMD}
\end{figure*}

We do not use group membership information from the group catalogues because our analysis is designed for a sample which includes the infalling galaxies around each group, as well as foreground and background `interlopers'. We therefore select satellite candidates from the SDSS catalogue within an aperture of $2.5\,r_{\rm vir}$, and $\pm 2\sigma_{\rm 3D}$ (we assume isotropic velocity distributions such that $\sqrt{3}\sigma_{\rm 1D} = \sigma_{\rm 3D}$). These apertures are large enough that essentially all galaxies which enter them and begin orbiting the central group never orbit back out of them \citep[e.g.][]{2013MNRAS.431.2307O}. For each satellite candidate, we determine normalized position and velocity offsets from the group centre as $R=d_{\rm A}\Delta\theta/r_{\rm vir}$ and $V=c|z_{\rm sat}-z_{\rm host}|/((1+z_{\rm host})\sigma_{\rm 3D})$, where $d_{\rm A}$ is the angular diameter distance, $z_{\rm sat}$ and $z_{\rm host}$ are the satellite and host redshifts, and $c$ is the speed of light. This results in a sample of $7.2\times10^4$ satellite candidates around $3.6\times10^3$ groups and clusters.

In order to assign galaxy halo mass estimates to observed satellites, we adopt the stellar-to-halo mass relation (SHMR) of \citet{2013ApJ...770...57B}. Our analysis, described below, does not rely on precision halo masses since the distribution of possible orbits for a halo with given $(R, V)$ is only a weak function of halo mass \citep{2013MNRAS.431.2307O} -- estimates within $\sim 0.5\,{\rm dex}$ should suffice. This SHMR matching is not directly applicable to satellites, which can be stripped of dark matter and/or stars. However, in the simulations we use the peak halo mass, which is still reasonably well estimated for satellites using the SHMR provided that the stellar component is not substantially stripped. As satellites heavily stripped of stars, but not yet completely destroyed, form only a small fraction of the satellite population \citep{2019MNRAS.485.2287B}, we do not attempt to account for these explicitly, and simply accept that their halo masses will be underestimated, introducing a weak bias in our analysis. Possible biases arising from the halo mass estimates are discussed further in Sec.~\ref{SubsubsecOLCompat}.

Finally, we use two diagnostics of star formation activity, as illustrated in Fig.~\ref{FigCMD}. The first uses the broad-band $(g-r)$ colour: we draw a line `by eye' just below the red sequence, defined as $(g-r)=0.05\log_{10}(M_\star/{\rm M}_\odot)+0.16$, and classify galaxies above (below) this line as `red' (`blue'). Since our analysis relies only on a binary separation of the two populations, our results are not strongly sensitive to the exact location of this cut, provided it reasonably separates the red and blue populations. The second diagnostic classifies galaxies as `active' or `passive' based on their specific SFR (sSFR), using the same limit as \citetalias{2016MNRAS.463.3083O}: ${\rm sSFR}/{\rm yr}^{-1}=-0.4\log_{10}(M_\star/{\rm M}_\odot)-6.6$. Our model parameters which describe the fractions of star-forming and quiescent galaxies inside/outside groups and clusters have a straightforward dependence on this choice -- for instance, moving the colour cut up in Fig.~\ref{FigCMD} would simply universally increase the fraction of blue galaxies. The more interesting parameters which describe the timing and timescale of the transition from red to blue, or active to passive, are insensitive to the location of the cut, within reason.

\begin{figure}
  \includegraphics[width=\columnwidth]{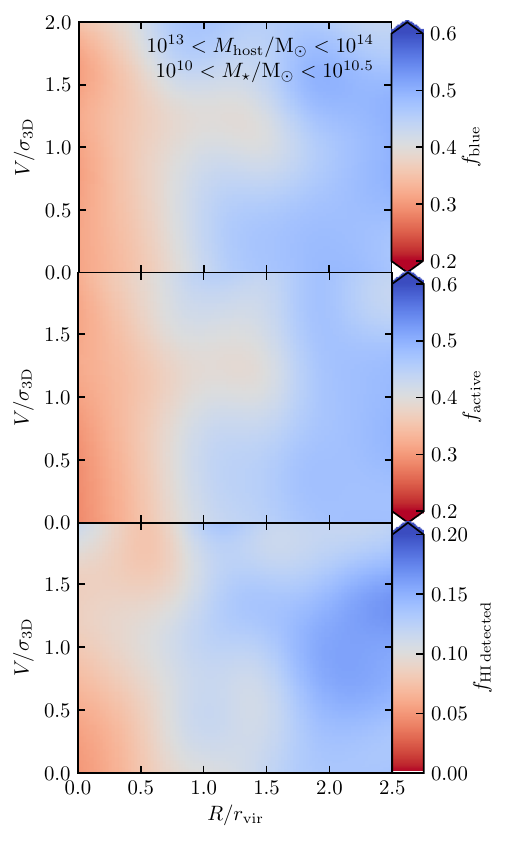}
  \caption{\emph{Upper panel:} Fraction of blue (see Fig.~\ref{FigCMD}) galaxies as a function of normalized projected position $R/r_{\rm vir}$ and velocity $V/\sigma_{3{\rm D}}$ offset from the host group centre, for galaxies with $10<\log_{10}(M_\star/{\rm M}_\odot)<10.5$ and $13<\log_{10}(M_{\rm host}/{\rm M}_\odot)<14$, of which there are $\sim 10^4$. There is a clear relative deficit of blue galaxies at low $R$ and $V$. We smooth the distributions of blue and red galaxies with a Gaussian kernel of width $0.25$ in both coordinates before computing the fraction to better highlight the overall trend. \emph{Middle panel:} As upper panel, but replacing the blue fraction with the `active fraction' (see Fig.~\ref{FigCMD}). \emph{Lower panel: }As above, but showing the fraction of \HI-detected galaxies; only those galaxies in the region where the ALFALFA and SDSS survey volumes overlap are included (see Sec.~\ref{SubsecDataStripping}), of which there are $6\times 10^3$. \label{FigDataIn}}
\end{figure}

We show a representative visualization of the input to our models in the upper and middle panels of Fig.~\ref{FigDataIn}, which illustrate the fraction of blue and active galaxies, respectively, as a function of position in the PPS plane. In this example we have selected galaxies with $10<\log_{10}(M_\star/{\rm M}_\odot)<10.5$ around hosts of $13<\log_{10}(M_{\rm host}/{\rm M}_\odot)<14$ (similar figures for more selections in $M_\star$ and $M_{\rm host}$ can be found in Appendix~\AppPPSMaps), and have smoothed the distributions of active and passive galaxies with a Gaussian kernel of width $0.25$ in each coordinate in order to bring out the overall trend: that there is a relative deficit of blue and active galaxies at low $R$ and $V$. This smoothing is for visualization only and does not enter into our analysis below.

\subsection{Sample for stripping analysis}
\label{SubsecDataStripping}

Our objective is to measure both the timing and timescale for gas stripping and star formation quenching, using a common sample of galaxies, and a common methodology. We could hope to measure, for instance, a delay, or absence thereof, between the removal of atomic hydrogen gas (e.g. by ram pressure) and the shutdown of star formation, which would provide qualitatively new constraints on the environmental quenching process at the galaxy population level.

We begin with the same galaxy sample as in Sec.~\ref{SubsecDataQuenching} but supplement it with \HI\ gas masses from the ALFALFA survey source catalogue \citep{2018ApJ...861...49H}. We match the optical counterparts in an extended version of that catalogue \citep{2011AJ....142..170H,2020arXiv201102588D} against the SDSS galaxy sample described above, requiring matches to be within $2\,{\rm arcsec}$ on the sky and with a redshift difference of less than $0.0005$ ($\sim 150\,{\rm km}\,{\rm s}^{-1}$; note that we are matching the positions of optical sources already associated to ALFALFA \HI\ detections, so these small tolerances are reasonable). The net result is a subset of the sample described in Sec.~\ref{SubsecDataQuenching}, occupying the overlap in sky and redshift coverage of the SDSS and ALFALFA surveys, with either an \HI\ mass measurement, or an \HI\ flux upper limit for non-detections. For non-detections, we derive approximate upper limits on the \HI\ mass assuming an inclination estimated from the $r$-band axis ratio reported in the SDSS catalogues ($i=\cos^{-1}(b/a)$), an (inclined) width $W_{50}^i$ for the \HI\ line estimated from the $g$-band Tully-Fisher relation of \citet[][table~3]{2017MNRAS.469.2387P}, Hubble flow distance estimates derived from the SDSS redshifts, and the ALFALFA 90~per~cent completeness limit in \HI\ flux $S_{21}$ as a function of $W_{50}$ \citep[][eq.~4]{2011AJ....142..170H}. The reduced survey volume results in a factor of $\sim2$ fewer galaxies, for a total sample of $3.7\times10^4$ satellite candidates within $R<2.5$ and $V<2.0$ of $3.1\times 10^3$ groups and clusters.

Analogous to the colour cut used to separate star-forming and quiescent galaxies (Sec.~\ref{SubsecDataQuenching}), we experiment with a variety of criteria to classify galaxies as `gas-rich' or `gas-poor'. However, the large fraction of \HI\ non-detections -- $90$~per~cent of SDSS galaxies in the region and redshift interval where the surveys overlap have no ALFALFA counterpart -- makes this challenging. Unlike the distribution of galaxies in colour-magnitude space, in $M_{\rm HI}$--$M_\star$ space there is no obvious separation into two populations, except perhaps the `detected' and `undetected' populations. This is intuitive: while a quiescent galaxy reddens but remains relatively easy to detect in an optical survey, a gas-poor galaxy becomes very challenging to detect in 21-cm emission. We have therefore experimented with divisions in $M_{\rm HI}$--$M_\star$ with various slopes -- e.g constant $M_\star/M_{\rm HI}$, constant $M_{\rm HI}$, intermediate slopes -- and normalizations. In each case we also consider different treatments of the $M_{\rm HI}$ upper limits, for instance: treating all upper limits as gas-poor; considering only upper limits which are constraining enough to discriminate between gas-rich and gas-poor, given a particular definition. We reliably find a gradient in the fraction of gas-rich galaxies as a function of position in PPS, with less gas-rich galaxies at low $R$ and $V$, independent of the definition of `gas-rich' used, within reason. None of the options explored being obviously superior to the others, we have opted to pursue our analysis using the simplest: we label `gas-rich' those galaxies which are detected in ALFALFA, and those not `gas-poor'\footnote{See Appendix~\AppAltGas\ for a demonstration that our main conclusions are robust against reasonable variations in this definition.}.

In the lower panel of Fig.~\ref{FigDataIn}, we show the resulting gas-rich fraction as a function of position in the PPS plane. A clear gradient is visible, such that there are fewer \HI-detected galaxies in groups and clusters, even though the fraction of `gas-rich' galaxies (i.e., detected in \HI) is globally much lower than would be expected for a deeper survey \citep[e.g.][]{2015ApJ...810..166E}. Because the information pertaining to the timing and timescale for gas stripping along a satellite orbit is encoded in the `shape' of the transition in the lower panel of Fig.~\ref{FigDataIn}, rather than in the absolute normalization of the distribution, we will be able to recover physically meaningful constraints on the relevant model parameters in our analysis below despite the numerous weak upper limits in the input catalogue. This depends crucially on the probability of a galaxy being detected in \HI\ (given its unknown \HI\ mass) being independent of its position in PPS. This is approximately true; we will return to a more detailed discussion of possible biases in Sec.~\ref{SecDiscussion}. Of course, an input catalogue based on a deeper survey would be preferable, however no other current surveys achieve sufficient depth covering a large enough volume for use in our statistical analysis. It seems probable, however, that this will change soon, as SKA precursor facilities come online and begin their surveys \citep[][and references therein]{2020Ap&SS.365..118K}; we plan to revisit our analysis as new data become available.

\section{Simulations}
\label{SecSims}

We make use of two simulation data sets: the Hydrangea cosmological hydrodynamical zoom-in simulations of clusters help to guide the form of our model for gas stripping and star formation quenching (Sec.~\ref{SubsecHydrangea}), and a periodic N-body volume run to a scale factor of $2$ (redshift $z=-\frac{1}{2}$) to allow us to infer the probability distributions for the orbits of observed galaxies in groups and clusters (Sec.~\ref{SubsecNbody}).

\subsection{Hydrangea}
\label{SubsecHydrangea}

The Hydrangea simulations are a suite of $24$ cosmological hydrodynamical `zoom-in' simulations. The zoom regions are selected around rich clusters ($M_{\rm vir}>10^{14}\,{\rm M}_\odot$) but extend out to $\sim 10r_{\rm vir}$ and so include many surrounding groups as well as field galaxies. The same galaxy formation model as in the EAGLE project \citep{2015MNRAS.446..521S,2015MNRAS.450.1937C} is used -- specifically the `AGNdT9' model -- at the same fiducial resolution level used for the $100\,{\rm Mpc}$ EAGLE simulation: a baryon particle mass of $1.81\times10^6\,{\rm M}_\odot$ and a force softening of $700\,{\rm pc}$ (physical) at $z<2.8$. Full details of the simulation setup and key results are described in \citet[][see also \citealp{2017MNRAS.471.1088B}]{2017MNRAS.470.4186B}, and we refer to the papers describing EAGLE, cited above, for details of the algorithms, models and calibration strategy. The EAGLE model does not explicitly model the neutral or atomic gas fractions of particles; we estimate atomic gas masses as described in \citet{2017MNRAS.464.4204C}, using the prescriptions from \citet{2013MNRAS.430.2427R} and \citet{2006ApJ...650..933B}.

The Hydrangea sample broadly reproduces many properties of galaxy clusters. Of particular relevance here is that the scaling with stellar mass (for $M_\star\gtrsim 10^{10}\,{\rm M}_\odot$) of the strength of the differential quenching effect due to the cluster environment is in quantitative agreement with observations by \citet{2012MNRAS.424..232W}, although the absolute quenched fraction is too low both in clusters and in the field \citep[see][their fig.~6]{2017MNRAS.470.4186B}. Since our model (Sec.~\ref{SecModel}) is explicitly designed to capture the differential effect of the cluster (or group) environment, these simulations are well-suited to offer guidance on its functional form.

Furthermore, Hydrangea offers a compromise between number of clusters ($\sim 40$ with $M_{\rm vir}>10^{14}\,{\rm M}_\odot$; several of the $24$ zoom regions contain additional clusters besides that centered in the volume) and resolution (we define `well-resolved' galaxies as those with $M_\star\geq2\times10^9\,{\rm M}_\star$, or about $10^3$ star particles) not found in other current cosmological hydrodynamical simulations. We stress, however, that the cold ISM is beyond the resolving power of these simulations: the cold ISM is not modelled, and a temperature floor is imposed, normalized at $8000\,{\rm K}$ for $n_{\rm H}=10^{-1}\,{\rm cm}^{-3}$, with the floor depending on density via an effective equation of state $P_{\rm eos}\propto\rho^{4/3}$. Gas in this regime follows empirically calibrated prescriptions for star formation and feedback. This means that dynamically cold, thin gas discs cannot exist in Hydrangea (see \citealp{2016MNRAS.456.1115B}, sec.~6.1, and \citealp{2018MNRAS.476.3648N}, sec.~3.5, for some additional details) -- the gas discs are therefore somewhat too weakly bound and are likely more susceptible to e.g. stripping by ram pressure than they should be. The Hydrangea cluster environment is therefore likely to strip satellites, especially low mass satellites, of gas and quench their star formation somewhat more efficiently than real clusters \citep[see ][fig.~6]{2017MNRAS.470.4186B}.

\subsection{N-body}
\label{SubsecNbody}

We broadly follow the methodology of \citetalias{2016MNRAS.463.3083O} to derive orbit parameter probability distributions from a library of orbits extracted from an N-body simulation, with some improvements.

We extend the `level 0' N-body simulation from the voids-in-void-in-voids \citep[VVV;][]{2020Natur.585...39W} project, using exactly the same configuration as for the original simulation except as described below. The simulation has a box size of $500\,h^{-1}\,{\rm Mpc}$, mass resolution elements of $10^9\,h^{-1}\,{\rm M}_\odot$, and a force softening of $4.6\,h^{-1}\,{\rm kpc}$, which offer a reasonable compromise between abundance of group- and cluster-sized structures and smallest resolved satellite galaxies. We run the simulation to a final scale factor of $a=2$ ($z=-\frac{1}{2}$, $\approx 10\,{\rm Gyr}$ into the future). This allows us to tabulate probability distributions for additional orbital parameters, in particular the time of first pericentre, even when it occurs in the future; from the distribution of pericentric times up to $a=2$, we estimate that $<0.1$~per~cent of $a=1$ satellites have not yet had a pericentric passage by $a=2$. We use the {\sc rockstar} halo finder \citep{2013ApJ...762..109B} and the related {\sc consistent-trees} utility \citep{2013ApJ...763...18B} to generate halo merger trees for all haloes with $>30$ particles ($M_{\rm vir}\gtrsim 4\times 10^{10}\,{\rm M}_\odot$, enough to resolve the $M_{\rm vir}\sim2.5\times 10^{11}\,{\rm M}_\odot$ haloes hosting the lowest stellar mass galaxies in our observed sample -- $M_\star=10^{9.5}\,{\rm M}_\odot$ -- until they have been stripped of $\gtrsim 85$~per~cent of their mass). We then identify satellites of host systems with $\log_{10}(M_{\rm vir}/{\rm M}_\odot)>12$ as those haloes within $2.5\,r_{\rm vir}$ at $z=0$. We trace the primary progenitors/descendants of the satellite sample backward/forward in time and record their orbits relative to the primary progenitor/descendant of their host system at $z=0$. We do not attempt to interpolate between simulation outputs, but instead simply adopt the output time immediately following an event as the time of that event. The output times are not uniformly spaced; the median time between outputs is $220\,{\rm Myr}$, and never exceeds $380\,{\rm Myr}$, sufficient to resolve the timescales which we measure in Sec.~\ref{SecResults}. We compile a table containing properties of the satellites at $z=0$: the projected offset from the cluster centre $R=\sqrt{(r_{{\rm host},x}-r_{{\rm sat},x})^2+(r_{{\rm host},y}-r_{{\rm sat},y})^2}/r_{\rm vir}$, the projected velocity offset $V=|(v_{{\rm host},z}-v_{{\rm sat},z})|+H(r_{{\rm host},z}-r_{{\rm sat},z})$, and the halo mass of the host, $M_{\rm host}$. $r$ and $v$ are the coordinate and velocity vectors of the simulated systems, with subscripts $(x,y,z)$ denoting the orthogonal axes of the simulation volume; $H$ is the Hubble parameter. For the satellite mass, $M_{\rm sat}$, we use the maximum mass at $z\geq 0$, which is better correlated with the stellar mass for moderately stripped satellites \citep[][see also appendix~A in \citealp{2013MNRAS.432..336W}]{2006ApJ...647..201C}. Finally, we also tabulate the lookback time to the first pericentric passage $t-t_{\rm fp}$ of the satellite within the $z=0$ host system, with negative times corresponding to future times ($a > 1$ or $z < 0$).

We also compile a similar sample of interlopers around each host system. These are haloes which are within $2.5\,r_{\rm vir}$ in projection (arbitrarily along the simulations $z$-axis), but outside $2.5\,r_{\rm vir}$ in $3{\rm D}$ -- foreground and background objects. We also require the line-of-sight ($z$-axis) velocities of interlopers to be within $\pm 2.0\,\sigma_{\rm 3D}$ of the host halo velocity along the same axis. We compile the values of $R$, $V$, $M_{\rm host}$ and $M_{\rm sat}$ for all interlopers.

\begin{figure*}
  \includegraphics[width=\textwidth]{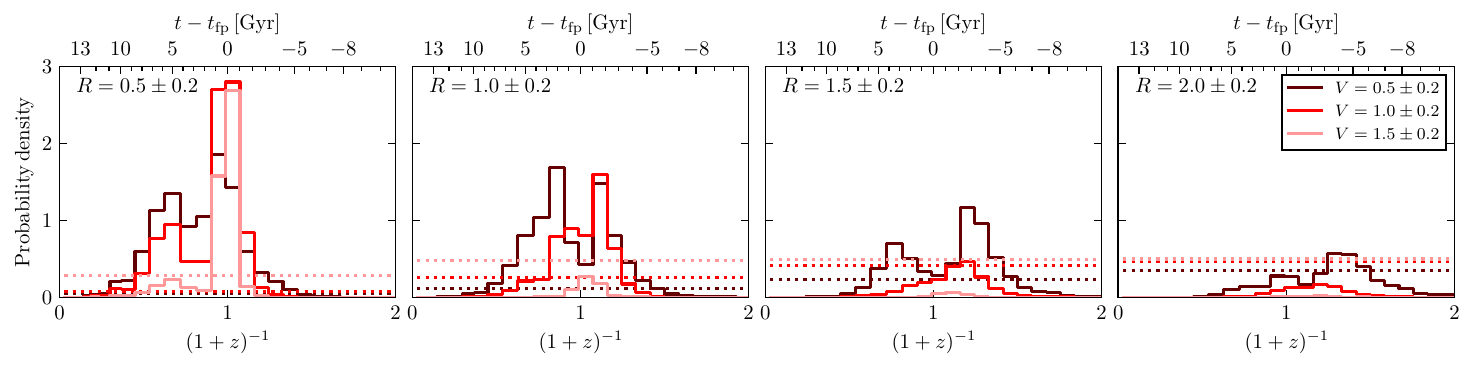}
  \caption{Sample probability distributions for the scale factor $(1+z)^{-1}$ of first pericentre or, equivalently, the time since first pericentre $t-t_{\rm fp}$. These distributions allow for the possibility that the first pericentric passage is in the future, in this case encoding information about how far in the future it will occur. All examples are for hosts with $10^{14}<M_{\rm host}/{\rm M}_\odot<10^{15}$ and satellites with $10^{11}<M_{\rm sat}/{\rm M}_\odot<10^{12}$. Each panel corresponds to a different radius $R$ (in units of $r_{\rm vir}$) from the host, as labelled on the panels, and different colours correspond to different velocity offsets $V$ (in units of $\sigma_{3{\rm D}}$), as labelled in the legend. The dotted lines illustrate the relative number of interlopers: the integrals of the solid and dotted curves over the plotted range are proportional to the number of satellites and interlopers, respectively.}
  \label{FigPDFDemo}
\end{figure*}

In order to estimate the probability distribution for the pericentric time for an observed satellite (or interloper) galaxy with a given ($R$, $V$, $M_{\rm host}$, $M_{\rm sat}$), we use the distribution of $t_{\rm fp}$ for all satellites and interlopers within ($0.05$, $0.04$, $0.5\,{\rm dex}$, $0.5\,{\rm dex}$) of each of these parameters, respectively. Our results are not sensitive to the exact intervals chosen for each parameter; we find that these values offer a good compromise between keeping a narrow range around the properties of the galaxy of interest and selecting a large enough subsample to construct a well-sampled probability distribution for $t_{\rm fp}$. The interlopers do not have measurements of $t_{\rm fp}$, but their abundance relative to the selected satellites defines the probability that the observed galaxy is an interloper rather than a satellite. In this way we compute probability distributions for $t_{\rm fp}$ individually tailored to each observed galaxy. We illustrate example $t_{\rm fp}$ probability distributions for satellites with $10^{11}<M_{\rm sat}/{\rm M}_\odot<10^{12}$ in hosts with $10^{14}<M_{\rm host}/{\rm M}_\odot<10^{15}$ at various locations in the PPS plane in Fig.~\ref{FigPDFDemo}.

\section{Environmental processing model}
\label{SecModel}

In this section we describe the statistical model which we use in Sec.~\ref{SecResults} to infer parameters describing quenching and stripping in groups and clusters. In Sec.~\ref{SubsecModelMotivation} we present results from the Hydrangea cosmological hydrodynamical simulations (see Sec.~\ref{SubsecHydrangea}) which motivate the form we adopt for our model. In Sec.~\ref{SubsecDefinitionFitting} we provide the formal definition of the model, and describe the method we use to constrain its parameters. Finally, in Sec.~\ref{SubsecModelTests} we describe two tests which demonstrate the limits within which our method can reliably recover the model parameters.

\subsection{Motivation}
\label{SubsecModelMotivation}

\begin{figure*}
  \includegraphics[width=\textwidth]{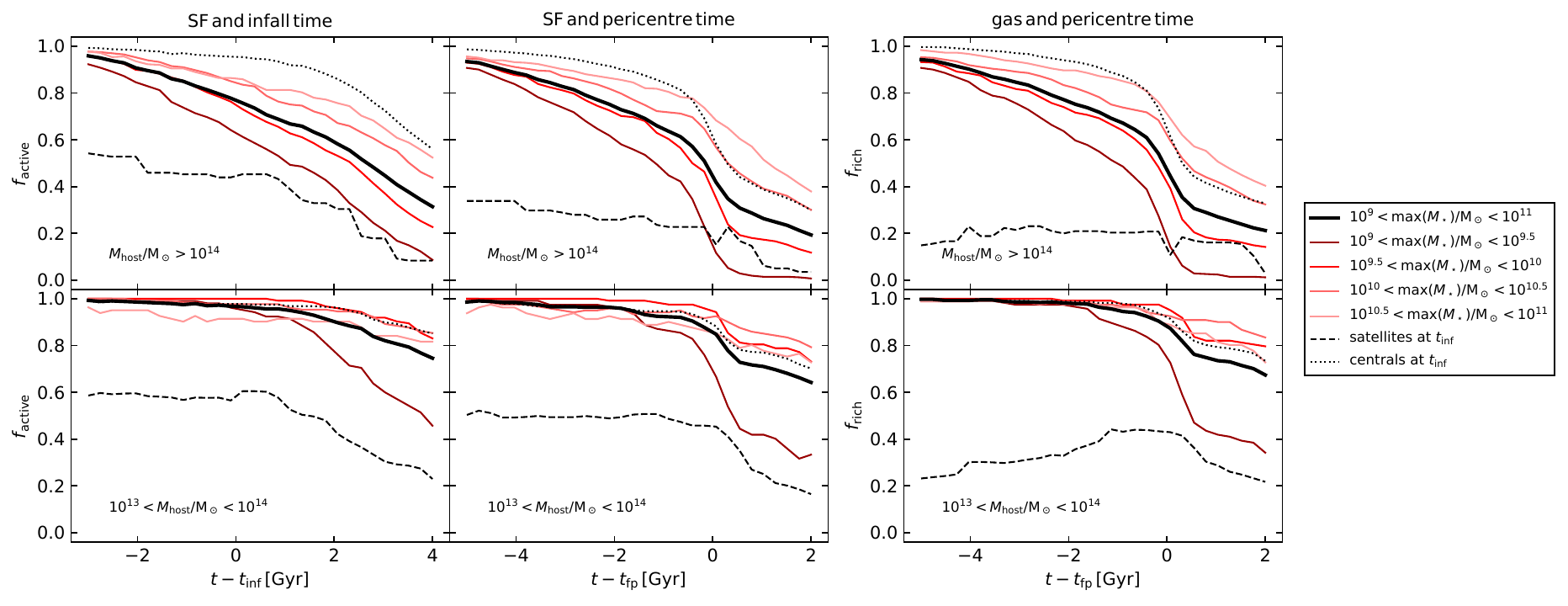}
  \caption{Fraction of star forming satellites, $f_{\rm active}$, defined as those with ${\rm sSFR} > 10^{-11}\,{\rm yr}^{-1}$ (columns 1 and 2), or of gas-rich satellites (column 3), $f_{\rm rich}$, defined as those with $M_{\rm HI}/M_\star>10^{-3}$, as a function of orbital time around Hydrangea clusters (upper row) and groups (lower row). In the first column, the orbital phase is aligned to the infall time $t_{\rm inf}$; in the second and third columns the reference time is $t_{\rm fp}$, the time of first pericentre. The heavy black line shows the trend for a fiducial sample. Coloured lines subdivide this sample by peak stellar mass, while broken lines subdivide it by those satellites which were centrals or satellites of another group at the time of infall.}
  \label{FigParamSelection}
\end{figure*}

We use results from the Hydrangea simulations to guide the form of our model linking the orbital histories of galaxies to their current star formation (or gas content). Inspired by previous studies \citep[\citetalias{2013MNRAS.432..336W,2016MNRAS.463.3083O};][]{2019MNRAS.488.5370L}, we first parametrized the orbital history by the infall time $t_{\rm inf}$, here defined as when the satellite first crosses $2.5\,r_{\rm vir}$. The left panels of Fig.~\ref{FigParamSelection} show the evolution of the active fraction (with `active' defined as ${\rm sSFR} > 10^{-11}\,{\rm yr}^{-1}$) as a function of the time since infall for the ensemble of Hydrangea satellites which fell into their hosts ($M_{\rm host}/{\rm M}_\odot>10^{14}$ in the upper panel, $10^{13} < M_{\rm host} / {\rm M}_\odot < 10^{14}$ in the lower panel) between $4$ and $10\,{\rm Gyr}$ ago\footnote{\label{FootnoteStacking}This selection allows us to track each individual galaxy in the sample across the entire $t-t_{\rm inf}$ range plotted without running into the beginning/end of the simulation. Note that this approach involves `shifting' the orbits of the satellites in time to align them on their infall or pericentre times; the results in Fig.~\ref{FigParamSelection} are therefore not representative of a fixed redshift. Part of the overall declining trend seen in all panels of Fig.~\ref{FigParamSelection} comes from the decline in the global fraction of star forming galaxies with time. An example `snapshot' at a fixed time is shown in Fig.~\ref{FigFitIllustrate}; see also similar figures in Appendix~\AppExtSim.} (heavy black line). The upper panels are for $\sim$cluster-mass hosts with $M_{\rm host} > 10^{14}\,{\rm M}_\odot$, while the lower panels are for $\sim$group-mass hosts with $10^{13}<M_{\rm host}/{\rm M}_\odot<10^{14}$. We see the expected monotonic decline in the active fraction as the population orbits for longer in the cluster. Perhaps surprisingly, the decline begins several Gyr before infall into the cluster-mass hosts. The reason for this becomes apparent when the galaxies are separated according to whether they were centrals (dotted line) or satellites (solid line) at infall: the early decline is predominantly driven by satellites, pointing to `pre-processing' in groups. The trend is also sensitive to the peak stellar mass of the satellites (coloured lines), with low-mass satellites (darker colour) feeling the influence of the host more strongly than high-mass satellites (paler colour).

In the middle panels of Fig.~\ref{FigParamSelection}, we repeat the same exercise as in the left panels, except that the orbits are aligned on the time of first pericentre, $t_{\rm fp}$, rather than the infall time. The same broad trends as in the left panels are seen, but a well-defined, sharp drop in $f_{\rm active}$ appears near $t-t_{\rm fp}=0$. This suggests that star formation quenching in Hydrangea is more tightly tied to the pericentric passage than initial infall into the group/cluster environment.

Finally, the right panels of Fig.~\ref{FigParamSelection} show the same as the central panel, except that the active fraction $f_{\rm active}$ has been replaced with the gas-rich fraction $f_{\rm rich}$, where gas-rich galaxies are defined as those with $M_{\rm HI}/M_\star>10^{-3}$. The close correspondance with the centre panels is striking -- in Hydrangea, satellites clearly experience substantial stripping near pericentre, often enough to immediately shut down star formation.

The behaviour illustrated in Fig.~\ref{FigParamSelection} is not unique to the Hydrangea simulations. \citet{2019MNRAS.483.5334S} find qualitatively, and approximately quantitatively, similar behaviour in the IllustrisTNG clusters. Their fig.~8 shows the same tendency for H\,{\sc i} stripping to be tightly tied to star formation quenching (e.g. the upper and centre panels of their figure are very similar). They also find that 50~per~cent (84~per~cent) of satellites (their satellite selection in IllustrisTNG has a similar stellar mass distribution to our selection in Hydrangea) are completely stripped/quenched after $2\,{\rm Gyr}$ ($3\,{\rm Gyr}$) in $M_{200}/{\rm M}_\odot>10^{14}$ hosts, and after $5\,{\rm Gyr}$ ($8\,{\rm Gyr}$) in $10^{13}<M_{200}/{\rm M}_\odot<10^{14}$ hosts. These values can be loosely compared to the times when the heavy black line crosses $f_{\rm active}=0.5$ ($0.16$) in Fig.~\ref{FigParamSelection}. We judge the two simulations to be in approximate agreement, though a precise comparison is hindered by the different halo mass definition, the different infall time definition -- our infall times should be earlier by approximately $2\,{\rm Gyr}$ -- and the limited window in infall times which we have used requiring some extrapolation to longer times since infall.

We draw our inspiration for a simple analytic model for quenching (or gas stripping) from previous work \citepalias{2016MNRAS.463.3083O} and from the results presented in Fig.~\ref{FigParamSelection}.

\subsection{Definition and fitting method}
\label{SubsecDefinitionFitting}

The model which we adopt relates the fraction of satellite galaxies which are actively star-forming\footnote{Or the fraction which are blue, or gas-rich; for brevity in this section we will use language which assumes an application to observations of sSFR.} to their time since first pericentre $t-t_{\rm fp}$. We explicitly handle galaxies which are still on their first approach and have not yet reached pericentre, these simply have a negative value of $t-t_{\rm fp}$. The model is intended to capture the relative effect of a host system on its satellites by comparing the properties of satellites of the host with the galaxy population immediately surrounding the host -- in practice any survey of a host also covers foreground/background galaxies which have projected positions and velocities consistent with the satellite populations. Our model therefore does not separately handle `pre-processing' in sub-groups falling into target hosts: members of such sub-groups contribute to the average properties of interlopers. We consider this a benefit rather than a drawback, as it means that we are sensitive only to the differential effect of the final host system. This formulation also ensures that our reference (`field') sample is exactly compatible with our satellite sample: a single selection on a parent catalogue yields both the reference and satellite populations together.

\begin{figure}
  \includegraphics[width=\columnwidth]{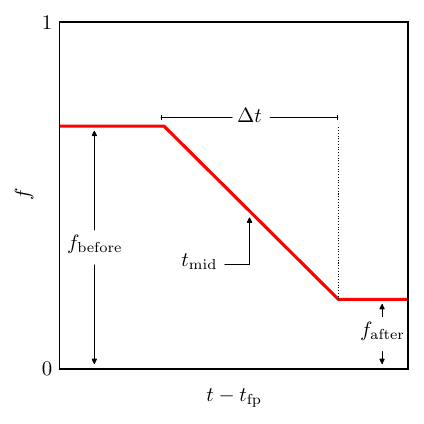}
  \caption{Schematic representation of the free parameters of the environmental processing model encoded in Eqs.~\ref{EqLinearDrop}--\ref{EqLnL}. The fraction $f$ of the galaxy population which is in the un-processed state (e.g. $f_{\rm blue}$, $f_{\rm active}$, $f_{{\rm HI}\,{\rm detected}}$) decreases linearly from $f_{\rm before}$ to $f_{\rm after}$ on a timescale $\Delta t$. The time when the drop is half complete is $t_{\rm mid}$. The reference time for a given galaxy is the time of its first pericentric approach to its final host, $t_{\rm fp}$.}
  \label{FigSchematic}
\end{figure}

The model has four free parameters: $(f_{\rm before}, f_{\rm after}, t_{\rm mid}, \Delta t)$. $f_{\rm before}$ and $f_{\rm after}$ describe the fraction of galaxies which are actively star forming before the effect of the host system begins to be felt, and after processing by the host is complete, respectively. We model the transition between these two states as a linear decline with time, with the reference time measured relative to the time of first pericentre $t_{\rm fp}$. $t_{\rm mid}$ sets the time at which half of the satellite population has been processed, while $\Delta t$ fixes the total width of the linear decline. These parameters are schematically illustrated in Fig.~\ref{FigSchematic}. When constraining model parameters, we adopt a flat prior probability distribution for each parameter, allowing values in the ranges: $0 < f_{\rm before} < 1$, $0 < f_{\rm after} < 1$, $-5 < t_{\rm mid}/{\rm Gyr} < 10$ and $0 < \Delta t/{\rm Gyr} < 10$. We also impose the constraint that $f_{\rm after} \leq f_{\rm before}$.

We constrain the parameters of our model by a maximum likelihood analysis. We perform independent analysis on independent sets of satellite galaxies, grouped by their stellar masses and host halo masses; our main results presented in Sec.~\ref{SecResults} are the parameter values as a function of $M_\star$ and $M_{\rm host}$. The redshift dependence of the parameters could also in principle be constrained, however for the purposes of this work we limit our analysis to low-redshift ($z<0.1$) satellites. We note a particularity of our appraoch: it is only sensitive to the quenching timescale for satellites of given $M_\star$, $M_{\rm host}$, and $t_{\rm fp}$ within $t_{\rm mid}\pm\Delta t$ of $t_{\rm fp}$. For instance, if the quenching timescale at early times was very short, a constant fraction $f_{\rm after}$ of satellites with ancient infall times will be observed to be passive -- there is no information contained in the measurements to actually constrain $t_{\rm mid}$ or $\Delta t$ for these galaxies. Put another way, $t_{\rm mid}$ is the typical time since (or until) $z\sim0$ satellites which are now being quenched had their first pericentric passage, which is conceptually distinct from the typical time to (or since) quenching for satellites having their first pericentric passage at $z\sim0$. A complete picture of quenching in dense hosts would therefore require a joint analysis of measurements across a range of redshifts.

The likelihood function ${\mathcal L}$ for our model is summarized as:

\begin{align}
  p_a(t-t_{\rm fp})&=
  \begin{cases}
    0 & {\rm if}\ t-t_{\rm fp} < t_{\rm mid}-\frac{\Delta t}{2}\\
    \frac{1}{2} + \frac{t - t_{\rm fp}-t_{\rm mid}}{\Delta t} & {\rm if}\ \left|t-t_{\rm fp}-t_{\rm mid}\right|\leq \frac{\Delta t}{2}\\
    1 & {\rm if}\ t-t_{\rm fp} > t_{\rm mid}+\frac{\Delta t}{2}
  \end{cases}\label{EqLinearDrop}\\
  p_{a,\,i} &= \frac{\int_{t=0}^{t_{\rm max}}p_a(t-t_{\rm fp})p_{{\rm peri},\,i}(R_i,V_i,t-t_{\rm fp}){\rm d}t}{\int_{t=0}^{t_{\rm max}}{\rm d}t}\label{EqIntegrals}\\
  p_{A,\,i} &= f_{\rm before} - p_{a,\,i}(f_{\rm before} - f_{\rm after})\label{EqFracs}\\
  P_i &=
  \begin{cases}
    p_{A,\,i} & {\rm if}\,{\rm active}(/\,{\rm blue}/\operatorname{gas-rich})\\
    1-p_{A,\,i} & {\rm if}\,{\rm passive}(/\,{\rm red}/\operatorname{gas-poor})
  \end{cases}\label{EqPi}\\
  \log{\mathcal L} &= \sum_i\log P_i\label{EqLnL}
\end{align}

Briefly, Eq.~\ref{EqLinearDrop} encodes our analytic model describing progress of the host in processing the satellites which it affects as a function of time, with a linear progression in the `processing fraction' $p_a$ from `none' before $t-t_{\rm fp}=t_{\rm mid}-\frac{1}{2}\Delta t$ to `all' after $t-t_{\rm fp}=t_{\rm mid}+\frac{1}{2}\Delta t$. Eq.~\ref{EqIntegrals} is the convolution of Eq.~\ref{EqLinearDrop} with the probability distribution for the pericentre time of the $i^{\rm th}$ satellite $p_{{\rm peri}, i}$, evaluated given its observable properties $(R_i, V_i)$. This weights the processing probability at a given time-to-pericentre by the probability that the satellite actually has that time-to-pericentre. In practice the integrals are evaluated as discrete sums, since the pericentre time probability distribution functions (see Sec.~\ref{SubsecNbody}) are discrete. Note that these probability distributions sum to $\leq 1$. In the case where the sum is less than one, the remaining probability budget corresponds to the probability that the `satellite' is in fact an interloper. Eq.~\ref{EqFracs} simply scales Eq.~\ref{EqIntegrals} by the fractions of active galaxies, $f_{\rm after}$ and $f_{\rm before}$. Eq.~\ref{EqPi} expresses that the probability for a given galaxy to appear in the sample is $p_{A,i}$ if that galaxy is active, or $1-p_{A,i}$ if it is passive (see Fig.~\ref{FigCMD}). Finally, Eq.~\ref{EqLnL} simply multiplies the probabilities for all galaxies, with the usual use of the logarithm to turn the product into a sum and keep the value of $\log{\mathcal L}$ within the realm of practical floating-point computation.

We estimate the posterior probability distribution of the model parameters by Markov chain Monte Carlo (MCMC) sampling using the likelihood function and priors described above, and the {\sc emcee} implementation \citep{2013PASP..125..306F} of the affine-invariant ensemble sampler for MCMC of \citet{2010CAMCS...5...65G}.

\subsection{Tests of the model}
\label{SubsecModelTests}

We perform two tests to check the accuracy of our model and fitting process. In the first, we use the same library of orbits drawn from our N-body simulation which was used to tabulate the probability distributions for $t-t_{\rm fp}$ (i.e. those illustrated in Fig.~\ref{FigPDFDemo}). We tabulate the projected phase space coordinates for each object in the library at $z=0$ (arbitrarily assuming a line of sight along the simulation $z$-axis). We assign each a stellar mass based on the \citet{2013ApJ...770...57B} SHMR, the same which we use (inverted) to assign halo masses to observed galaxies. We then choose fiducial values for the model parameters $(f_{\rm before}, f_{\rm after}, t_{\rm mid}, \Delta t)$ and randomly flag each object as active or quenched with a probability defined by the model parameters and the objects $t-t_{\rm fp}$ at $z=0$ as determined from its orbit. In this way we obtain a sample of data which is exactly described by the model which we wish to fit, and additionally has a distribution of orbits exactly consistent with those which will be used to infer $t-t_{\rm fp}$ from the projected phase space coordinates. This is therefore a `best case scenario' data sample.

From this sample, we draw a random subsample of $2000$ objects (in a narrow range of stellar mass) and draw a MCMC sample of the posterior distribution for the model parameters as described in Sec.~\ref{SubsecDefinitionFitting}. We repeat this exercise $5000$ times for different random subsamples and find that we recover an unbiased estimate of the input model parameters $f_{\rm before}$, $f_{\rm after}$ and $t_{\rm mid}$. The parameter $\Delta t$, however, tends to be underestimated, with a probability density peaking at $0$ even when the input $\Delta t$ is $>0$, and a median value typically underestimated by up to $\sim 1\,{\rm Gyr}$. However, we find that in all cases the confidence intervals are representative of the uncertainties. The true parameter values are without exception within the $68$, $95$ and $99$~per~cent confidence intervals of the estimates at least $68$, $95$ and $99$~per~cent of the time (sometimes slightly more, suggesting that the widths of the confidence intervals are modestly overestimated). We also repeat this exercise with smaller subsamples, down to a minimum of $100$ objects. We find that the confidence intervals, though wider, continue to accurately represent the uncertainty in the estimates.

\begin{figure}
  \includegraphics[width=\columnwidth]{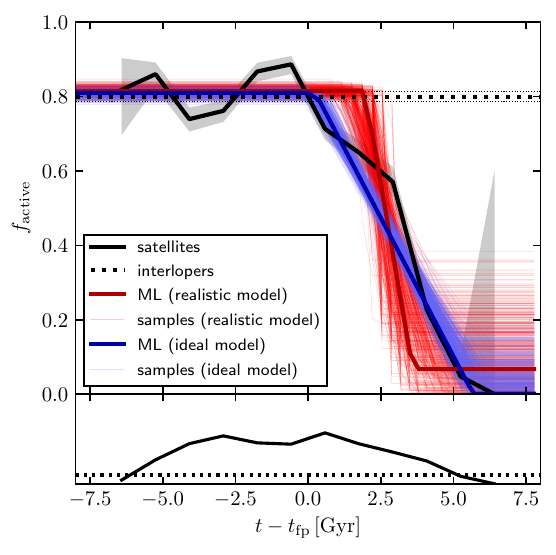}
  \caption{Illustration of our model constrained by Hydrangea data at $z=0.64$, where the solution is known, for the stellar mass bin centered at $M_\star\sim5\times10^{10}\,{\rm M}_\odot$ (see Fig.~\ref{FigModelTest}). The active fraction $f_{\rm active}$ as a function of time to first pericentre $t-t_{\rm fp}$, calculated in $0.5\,{\rm dex}$ bins, is shown with the black solid line -- the shaded band marks the $1\sigma$ confidence interval, estimated as proposed in \citet{1986ApJ...303..336G}. The horizontal dashed lines mark the active fraction and $1\sigma$ confidence interval for the interloper population. The solid line in the lower panel illustrates the relative counts in each bin; the dotted line is for the interlopers, normalized such that the integrals of the two curves are proportional to the relative abundance of interlopers and satellites. The blue curves are individual samples from the Markov chain computed using a model which is given the exact value of $t-t_{\rm fp}$ for each satellite, while the red curves are similar but for a model where $t-t_{\rm fp}$ is estimated from the observed location in phase space of each satellite (see Sec.~\ref{SubsecModelTests} for details). The heavier lines of each colour mark the sample from the chain with the highest likelihood.}
  \label{FigFitIllustrate}
\end{figure}

\begin{figure*}
  \includegraphics[width=\textwidth]{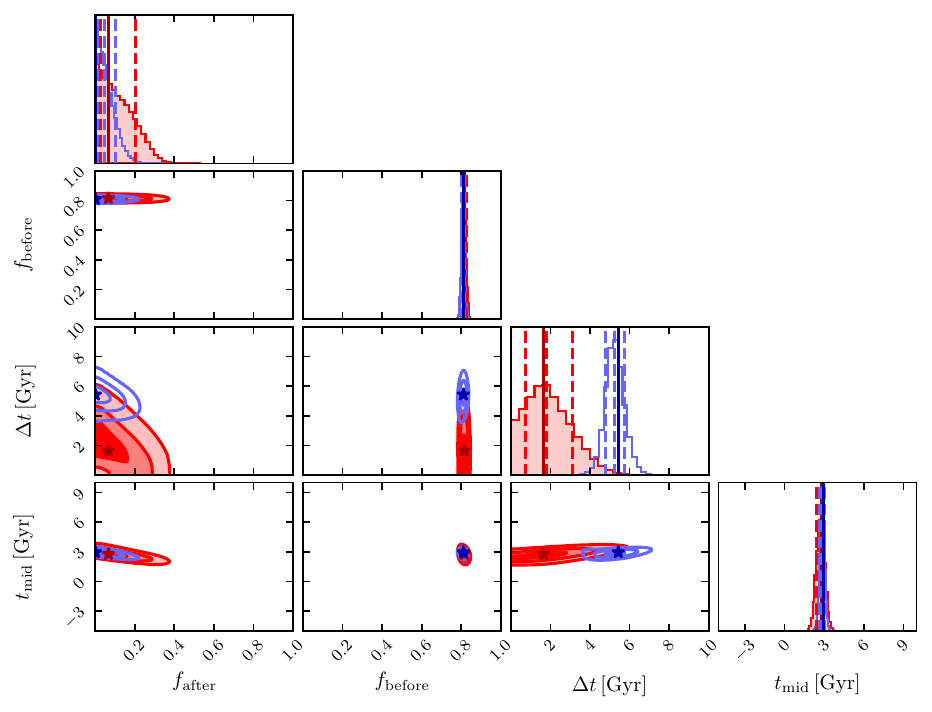}
  \caption{Example one- and two-dimensional marginalized posterior probability distributions for the parameters of the model defined in Sec.~\ref{SecModel} constrained by mock sSFR data from the Hydrangea simulations at $z\sim0.64$. The example shown here corresponds to the stellar mass bin at $\sim5\times10^{10}\,{\rm M}_\odot$ in Fig.~\ref{FigModelTest}. Open blue contours/histograms correspond to a fit where full knowledge of $t-t_{\rm fp}$ for each object is provided to the model (the `truth'), while filled red contours/histograms correspond to fits where $t-t_{\rm fp}$ is estimated from projected phase space coordinates. Contours are drawn at 68, 95 and 99~per~cent confidence intervals, and dashed lines are drawn at the $16^{\rm th}$, $50^{\rm th}$ and $84^{\rm th}$ percentiles. The stars and solid lines mark the position of the maximum likelihood parameter sample drawn in the Markov chain. Similar figures for all fits presented in Figs.~\ref{FigModelTest} and \ref{FigObsFits} are included in the Appendices~\AppExtSim\ and \AppCorners.}
  \label{FigCornerExample}
\end{figure*}

The second test which we perform uses mock data drawn from the Hydrangea simulations. In order to allow for a scenario in which the value of $t-t_{\rm fp}$ is known for each object in the sample, we make our mock observations on the $z\sim0.64$ snapshot of the simulation (approximately the midpoint in lookback time) such that we can track the orbits forward for objects with negative $t-t_{\rm fp}$. Again, we arbitrarily choose the simulation $z$-axis as the line of sight and tabulate the projected phase space coordinates of satellites within $R<2.5$ and $V<2.0$ around each cluster with $M_{\rm host}>10^{14}\,{\rm M}_\odot$, i.e. including interlopers. We estimate pre-infall halo masses from the stellar masses using the SHMR\footnote{We also repeated the same test using the exact maximum virial masses from any time $z\geq0.64$ of the satellites and found no significant change in the parameter estimates.} for $z=0.64$ of \citet{2013ApJ...770...57B}. Objects with ${\rm sSFR}>10^{-11}\,{\rm yr}^{-1}$ are flagged as active, and those below this threshold as quenched. This definition differs from those used for observed galaxies, but we note (i) that all we require for our model is a binary split of the galaxy population, so simply assuming that the bimodal ${\rm sSFR}$ distribution in the simulations and the observed bimodal $(g-r)$ colour and sSFR distributions broadly reflect the same active/passive populations seems reasonable, and (ii) we do not attempt to draw detailed comparisons between the parameters estimated for the simulations and those estimated for the observations, rather using the simulations only as a test for our methodology (but see Sec.~\ref{SecConc} for some discussion of our results in the context of recent simulations, including Hydrangea). For consistency, we also compute new $t-t_{\rm fp}$ probability distributions at $z=0.64$ from our N-body simulation. We exclude poorly resolved galaxies, retaining only those with $M_\star>2\times10^9\,{\rm M}_\odot$ ($\gtrsim10^3$ star particles). This leaves $\sim 1.1\times10^4$ satellites and interlopers.

\begin{figure}
  \includegraphics[width=\columnwidth]{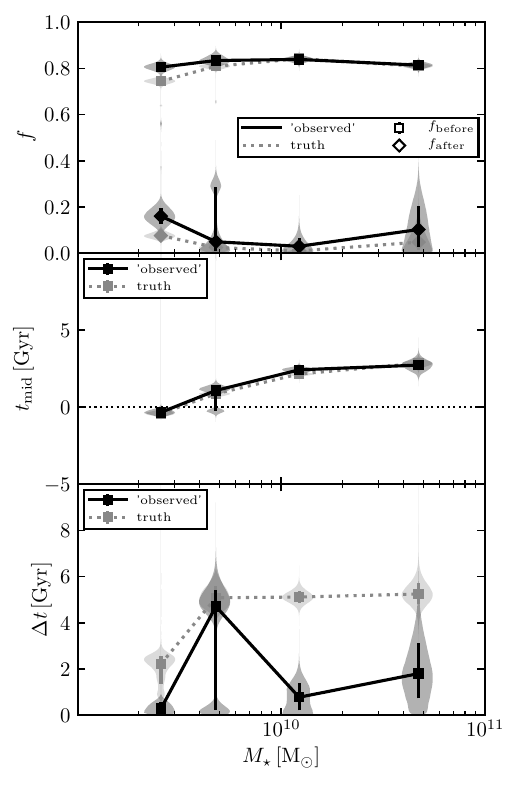}
  \caption{The marginalized median and $16^{\rm th}$--$84^{\rm th}$ percentile confidence interval for the parameters of the model defined in Sec.~\ref{SecModel} constrained by mock sSFR data from the Hydrangea simulations at $z\sim0.64$ are shown with points and error bars; the transparent `violins' show the full marginalized posterior probability distribution for each parameter. The lighter symbols of each type correspond to fits where full knowledge of $t-t_{\rm fp}$ for each object is provided to the model (the `truth'), while darker symbols correspond to fits where $t-t_{\rm fp}$ is estimated from projected phase space coordinates. The upper panel shows the active fractions before ($f_{\rm before}$, circles) and after ($f_{\rm after}$, squares) quenching by the host. The centre panel shows the quenching timing parameter $t_{\rm mid}$, and the lower panel the quenching timescale $\Delta t$. The sample is selected to have $M_\star>2\times10^9\,{\rm M}_\odot$ and is split into 4 stellar mass bins with equal counts. The symbols are plotted at the median stellar mass in each bin, offset by a small amount to ensure that the error bars are legible.}
  \label{FigModelTest}
\end{figure}

We rank the simulated galaxies by stellar mass and split the sample into 4 bins with even counts. For each of these independent subsamples, we estimate the parameter values of the model described in Sec.~\ref{SubsecDefinitionFitting} by MCMC sampling the posterior probability distribution, thus inferring the stellar mass dependence of the model parameters. In each case, we run two fits. In the first, we replace the probability distribution for $t-t_{\rm fp}$ for each object with the exact value as determined by tracking its orbit (we also inform the model as to which `satellites' are actually interlopers). This allows us to quantify the behavior of the model given perfect knowledge of the orbits. In Fig.~\ref{FigFitIllustrate}, we compare the distribution of model realizations from this Markov chain (blue lines) with the true active fraction as a function of $t-t_{\rm fp}$ (black line; this is related to, but not the same as, the curves shown in Fig.~\ref{FigParamSelection}, see footnote~\ref{FootnoteStacking} above) for the stellar mass bin centered at $M_\star\sim5\times10^{10}\,{\rm M}_\odot$ (see Fig.~\ref{FigModelTest}), demonstrating that our model achieves a good description of the underlying data when given optimal information. We also show the one- and two-dimensional marginalized posterior probability distributions for the $4$ model parameters with open blue histograms/contours in Fig.~\ref{FigCornerExample}. Figures similar to Figs.~\ref{FigFitIllustrate}~and~\ref{FigCornerExample} for the other stellar mass bins are included in Appendices~\AppExtSim\ and \AppCorners. In all cases, we find that this model fit is a fair representation of the underlying data, or `truth'. The parameter constraints as a function of stellar mass are summarized in Fig.~\ref{FigModelTest}, drawn with the lighter tone of each colour/symbol type, and dotted lines.

We fit the model a second time in each stellar mass bin, this time using the probability distributions for $t-t_{\rm fp}$ to infer this quantity based on the `observable' properties of the satellites/interlopers. This represents treating the simulations as closely as possible as an observed data sample. The corresponding model realizations are shown with red lines in Fig.~\ref{FigFitIllustrate} and filled red contours/histograms in Fig.~\ref{FigCornerExample}. The parameter constraints as a function of stellar mass are shown with the darker symbols and solid lines in Fig.~\ref{FigModelTest}. As in the first test described above, we find that the quenching timescale $\Delta t$ is systematically underestimated (the solid line in the lower panel of Fig.~\ref{FigModelTest} lies well below the dotted line). In contrast with the test using the model `painted onto' the N-body orbit library, however, this time the `true' values fall significantly outside the 68~per~cent confidence intervals in all cases. The timing of quenching, i.e. $t_{\rm mid}$, is still well-recovered (centre panel of Fig.~\ref{FigModelTest}). The active fractions $f_{\rm before}$ and $f_{\rm after}$ are generally well-recovered. We have been unable to identify the origin of the slight but formally significant overestimates at low $M_\star$.
Encouragingly, this does not seem to impact the accurate recovery of $t_{\rm mid}$, and does not seem to be related to the bias in $\Delta t$ (e.g. there is no visible degeneracy between $f_{\rm before}$ and $\Delta t$ in Fig.~\ref{FigCornerExample} or similar figures in Appendix~\AppExtSim).

We note the presence of two `peaks' of significant probability density for the parameters corresponding to the second ($M_\star\sim5\times10^{9}\,{\rm M}_\odot$) stellar mass bin, visible as two separate bulges in the `violins' in Fig.~\ref{FigModelTest}. Inspecting the pairwise marginalized probability distributions for the parameters (Appendix~\AppExtSim, Fig.~\AppExtSim{}6), we find that the lower $f_{\rm after}$ peak is associated to the higher $t_{\rm mid}$ peak, and the higher $\Delta t$ peak. Such degenerate solutions occured occasionally while we experimented with tests of our model, and we found that the lower $f_{\rm after}$ peak (usually consistent with $f_{\rm after}=0$) invariably corresponded to the `true' parameter values. This observation will be used in Sec.~\ref{SecResults} to motivate fixing $f_{\rm after}=0$ in our fiducial parameter estimates.

Together, these two tests suggest that when we apply the same method to observational data, our estimates of $f_{\rm before}$, $f_{\rm after}$ and $t_{\rm mid}$ are likely to be accurate. $\Delta t$ seems to be much more difficult to constrain -- the results to the two tests considered together suggest that constraints for this parameter should be taken as lower limits, which will motivate our treatment of $\Delta t$ as a `nuisance parameter' in Sec.~\ref{SecResults}.

\section{Characteristic timing of quenching and stripping}
\label{SecResults}

We now turn to the constraints on our model parameters using the observational inputs from SDSS and ALFALFA (Secs.~\ref{SubsecDataQuenching} and \ref{SubsecDataStripping}). After experimenting with these data, we have made two choices in the presentation of our fiducial results in this section. The first is to omit the parameter $\Delta t$ from the discussion in this section. The probability distributions for this parameter tend to be very broad and, given the biases seen in tests in Sec.~\ref{SubsecModelTests}, are difficult to interpret. We still allow this parameter to vary with the same prior described above (flat in the interval $0$--$10\,{\rm Gyr}$), but treat it as a nuisance parameter which we marginalize over in the discussion below. Details on the constraints for $\Delta t$ are, however, included in Appendix~\AppExtRes.

The second fiducial choice which we make is to fix the parameter $f_{\rm after}$ to $0$. In the majority of cases this is the preferred value when the parameter is left free in any case\footnote{In cases where it is not the preferred value, there are usually multiple peaks -- it is likely that the $f_{\rm after}\sim0$ peak corresponds to the more correct set of parameter estimates, as discussed in Sec.~\ref{SubsecModelTests}.}. Furthermore, while $f_{\rm after}>0$ is mathematically straightforward, its physical interpretation is not. In principle is represents the fraction of galaxies which are blue/active/gas-rich once the group/cluster environment has had `long enough' to exert its influence, where `long enough' is encoded in $\Delta t$. This leads to a degeneracy between the two parameters: if one waits longer (higher $\Delta t$), more galaxies are quenched/stripped (lower $f_{\rm after}$) -- this is visible in the $\Delta t$ vs. $f_{\rm after}$ panel of Fig.~\ref{FigCornerExample}. This, in conjunction with $\Delta t$ being poorly constrained as explained above, leaves the interpretation of $f_{\rm after}$ somewhat ambiguous as well. Finally, we note that allowing $f_{\rm after}$ to vary does not change our qualitative conclusions, and makes only small quantitative differences. Neither $t_{\rm mid}$ nor $f_{\rm before}$ exhibit any apparent degeneracy with $f_{\rm after}$ (see Fig.~\ref{FigCornerExample}, and Appendices~\AppExtSim\ and \AppCorners). For completeness, the probability distributions including $f_{\rm after}$ as a free parameter are included in Appendix~\AppExtRes.

Lastly, before moving on to the actual parameter constraints, we re-iterate the physical interpretation of the two parameters which are the focus of our discussion below:
\begin{itemize}
\item $f_{\rm before}$ is the fraction of blue/active/gas-rich galaxies `outside' the cluster. This is determined by the combination of the galaxies which are in the group/cluster but have not yet felt its effects, and those interlopers within $R<2.5$ and $V<2.0$. These relatively small apertures around the hosts mean that $f_{\rm before}$ is a measure of the blue/active/gas-rich fraction \emph{just} outside the hosts, i.e. including pre-processed galaxies. Our measurement is therefore sensitive to the \emph{differential} effect of the final host, rather than the cumulative effect of all hosts for galaxies which fall into larger hosts while already being members of smaller groups.
\item $t_{\rm mid}$ is the characteristic time along their orbits when galaxies transition from blue/active/gas-rich to red/passive/gas-poor. Put another way, if a randomly selected blue/active/gas-rich satellite is dropped into a cluster, at time $t_{\rm mid}$ (recalling that $t_{\rm mid}=0$ corresponds to the time of first pericentre) there is a 50~per~cent chance that it has become red/passive/gas-poor. Put yet another way, assuming the blue-red/active-passive/gas-rich-to-poor transition for an \emph{individual} galaxy is rapid -- motivated by the bimodal distributions\footnote{It is less clear that the gas-fraction distribution is bimodal, motivating some caution when discussing the stripping $t_{\rm mid}$, below.} of Fig.~\ref{FigCMD} -- $t_{\rm mid}$ is a measure of the typical time within the population when this rapid transition occurs.
\end{itemize}

\begin{figure*}
  \includegraphics[width=\textwidth]{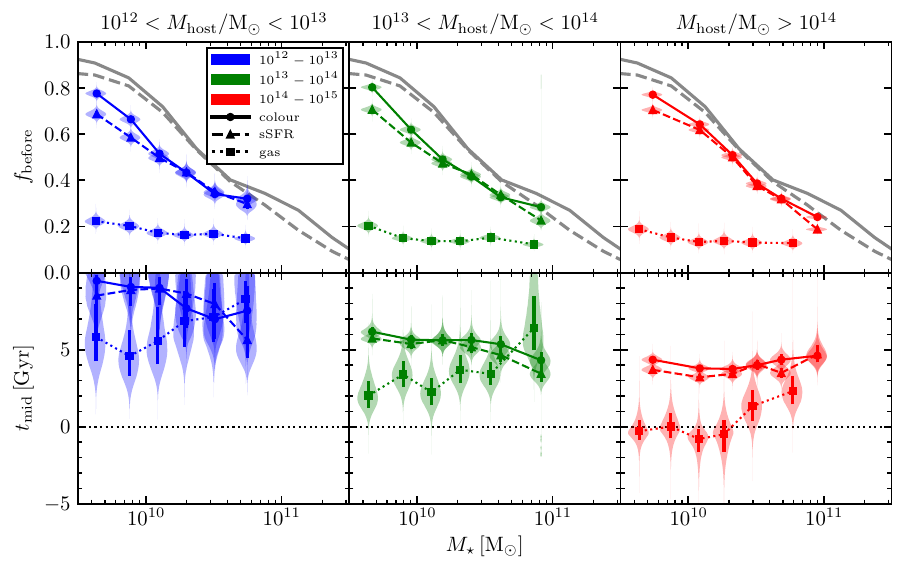}
  \vspace{-.7cm}\\
  \includegraphics[width=\textwidth]{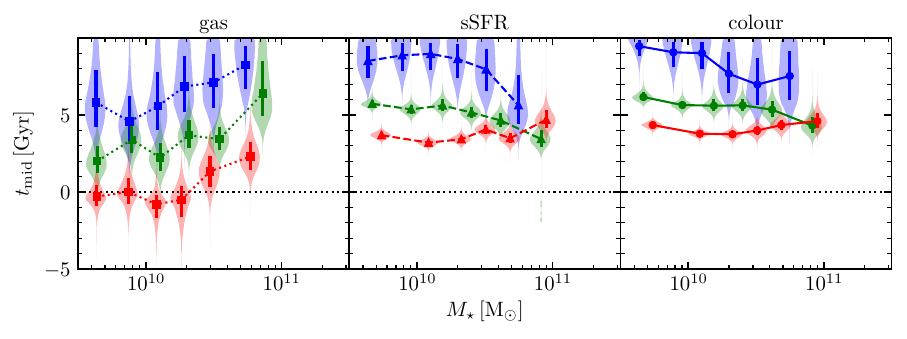}
  \caption{\emph{Upper panels}: Marginalized posterior probability distributions for the $f_{\rm before}$ (upper row) and $t_{\rm mid}$ (lower row) model parameters as a function of stellar mass around hosts of $10^{12}<M_{\rm host}/{\rm M}_\odot<10^{13}$ (left column, blue), $10^{13}<M_{\rm host}/{\rm M}_\odot<10^{14}$ (centre column, green), and $M_{\rm host}/{\rm M}_\odot>10^{14}$ (right column, red). The parameters are estimated for two tracers of star formation quenching -- $(g-r)$ colour (circles connected with solid lines) and ${\rm sSFR}$ derived from Balmer emission line strength (triangles connected with dashed lines) -- and for gas stripping as traced by detection in the ALFALFA survey (squares connected by dotted lines). Points mark the median value of each probability distribution, and error bars the $16$--$84^{\rm th}$ percentile confidence intervals; the transparent `violins' show the full marginalized posterior probability distributions. (The parameter estimates corresponding to the rightmost blue and green squares are likely spurious, see Sec.~\ref{SubsubsecStatConsid}.) The gray solid (dashed) line shows the overall blue (active) fraction of galaxies in the parent SDSS sample in the redshift interval $0.01 < z < 0.1$. \emph{Lower panels}: Exactly as the lower row in the upper panels, but re-organized to highlight trends with host mass: each column is for a single tracer (from left to right: \HI, Balmer emission lines, broadband colour), rather than a single host mass interval.}
  \label{FigObsFits}
\end{figure*}

An individual, independent Markov chain of model parameters is evaluated for each combination of input galaxy sample properties: host mass ($3$ bins: $10^{12}-10^{13}$, $10^{13}-10^{14}$, $>10^{14}\,{\rm M}_\odot$), satellite stellar mass (galaxies are ranked by $M_\star$ and separated into $6$ bins with equal counts), and tracer property ($(g-r)$ colour, sSFR, or \HI\ content). 

In the upper panels of Fig.~\ref{FigObsFits}, we show the marginalized posterior probability distributions for $f_{\rm before}$ and $t_{\rm mid}$ derived from each of these chains, focusing on the differences between the different tracer properties. $f_{\rm before}$ is a declining function of stellar mass for both the blue fraction (circles connected by solid lines) and the active fraction (triangles connected by dashed lines). This trend mirrors that of the overall galaxy population in SDSS -- overwhelmingly composed of `central' galaxies -- plotted with the solid (blue fraction) and dashed (active fraction) gray lines, but is offset to lower values by $\sim 0.05-0.2$. This highlights the importance of `pre-processing': the ensemble of interlopers and satellites just entering their hosts does not resemble the global average galaxy population. As a result of their evolution in a denser-than-average environment, interlopers and satellites just entering their current host systems are less likely to be blue and star forming.

The gas-rich fraction (squares connected by dotted lines) are much lower and flatter -- due to the limited depth of the ALFALFA survey these are certainly underestimates (see Sec.~\ref{SubsecDataStripping}), but this incompleteness is not expected to bias the corresponding estimates of $t_{\rm mid}$ (see Sec.~\ref{SecDiscussion} for further details).

We note that the input sample of galaxies for the gas analysis is a subset of that used for the colour and sSFR analyses, corresponding to the overlap region between the ALFALFA and SDSS surveys in redshift (see Fig.~\ref{FigObsOverview}) and sky coverage. This is the reason for the horizontal offset between the square symbols and the triangles and circles in Fig.~\ref{FigObsFits}. We have verified that using exactly the same input galaxies for all three analyses does not change the results of the colour and sSFR analyses, other than somewhat widening the confidence intervals.

\subsection{Quenching lags stripping}
\label{SubsecResultsA}

The central result of our analysis is shown in the second row of Fig.~\ref{FigObsFits}. The characteristic time $t_{\rm mid}$ when galaxies transition from blue to red within their host (circles connected by solid lines) is consistently found to be well after the first pericentric passage, by $\sim 4$--$5\,{\rm Gyr}$ in the highest mass hosts (lower right panel) up to perhaps $\sim7$--$9\,{\rm Gyr}$ in lower mass hosts (lower left panel), although here the confidence intervals are somewhat wider. The characteristic time when star formation activity ceases (as traced by the disappearance of Balmer emission lines) is coincident with or slightly (a few hundred ${\rm Myr}$ to a ${\rm Gyr}$) earlier than the colour transition. Such a short delay is not unexpected as the time taken for the stellar population to age and redden once star formation ceases is somewhat longer than that for the emission lines to disappear \citep[by about $\sim 300\,{\rm Myr}$, e.g.][]{2004MNRAS.348.1355B}. The characteristic time when galaxies are stripped of neutral hydrogen, on the other hand, is well before the quenching time (whether traced by sSFR or colour)\footnote{More properly, when they are sufficiently stripped to fall below the ALFALFA detection threshold. We note that, when including only galaxies below the ALFALFA redshift limit of $z=0.06$, the redshift distributions of satellites around low-, intermediate- and high-mass hosts are reasonably similar (see Fig.~\ref{FigObsOverview}), so a bias in distance is unlikely to be driving this result. We also note that the qualitative statement that `quenching lags stripping' is robust to reasonable changes in the definition of a `gas rich' galaxy, see Appendix~\AppAltGas.}, by $2$--$5\,{\rm Gyr}$. In the most massive hosts (upper panels, right column of Fig.~\ref{FigObsFits}), stripping seems to be well underway even $\gtrsim 1\,{\rm Gyr}$ before the first pericentric passage, while around lower mass hosts (centre and left columns) satellites appear to keep the bulk of their \HI\ until up to several ${\rm Gyr}$ after the first pericentric passage. We regard this difference between the $t_{\rm mid}$ values for star formation quenching (traced by colour or emission lines) and neutral gas stripping as strong evidence for continued star formation well after the onset of ram-pressure stripping of \HI. This is consistent with a `starvation' quenching scenario, although from our measurements we cannot discriminate between the molecular gas directly fueling star formation eventually being depleted, or alternatively being stripped on a subsequent pericentric passage. We will discuss this interpretation further in Sec.~\ref{SubsecInterpret} below.

We repeat the same information shown in the second row of Fig.~\ref{FigObsFits} in the lower panels, but re-arrange the curves to highlight differences between hosts of different masses. There is a clear trend for satellites of a given stellar mass to be stripped (left), become passive (centre), and be redden (right) earlier around more massive hosts, reflecting the generally harsher nature of higher density environments.

\section{Discussion}
\label{SecDiscussion}

We first consider the reliability of our results in the context of various statistical and systematic effects (Sec.~\ref{SubsecRobust}) before comparing with the results of other studies (Sec.~\ref{SubsecCompare}), and discussing the inferences that can be drawn regarding the processes governing the evolution of satellite galaxies based on our measurements (Sec.~\ref{SubsecInterpret}).

\subsection{Robustness of parameter constraints}
\label{SubsecRobust}

\subsubsection{Completeness of input catalogues}
\label{SubsubsecCompleteness}

We first consider the various biases and systematic effects which could influence the parameter estimates presented in Sec.~\ref{SecResults}.

The main systematic biases which are of concern for our statistical analysis are any which cause galaxies with a given property to be preferentially included in our sample. Biases which are tied to the PPS coordinates are of particular concern as these can affect the timescale $t_{\rm mid}$; PPS-independent biases will primarily affect $f_{\rm after}$ and $f_{\rm before}$. We consider as an example a single Markov chain, corresponding to a given interval in $M_{\rm host}$ and $M_\star$. We now suppose that in the input galaxy sample red galaxies are preferentially included relative to blue ones. If this occurs uniformly across the $(R, V)$ PPS plane, the result will be a lowering of the estimates for $f_{\rm before}$ and $f_{\rm after}$. The $t_{\rm mid}$ parameter, on the other hand, is tied to how long each satellite has been orbiting its host, and information about orbital phase comes exclusively from the PPS coordinates, so $t_{\rm mid}$ can only be affected by a bias if it is not uniform across PPS.

\begin{figure}
  \includegraphics[width=\columnwidth]{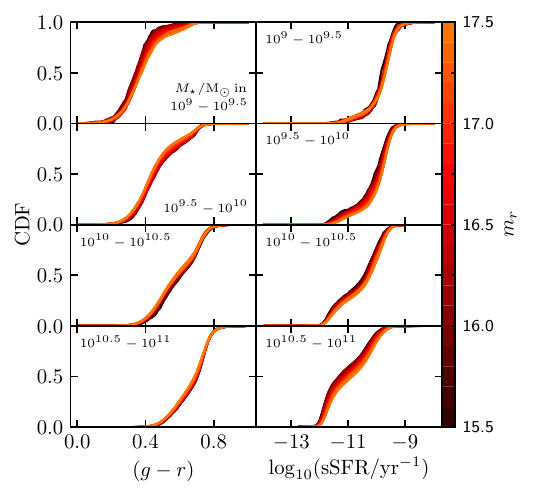}
  \caption{Assessment of colour and sSFR biases in the SDSS input catalogue. The cumulative distribution of galaxy colours (left column) and sSFRs (right column) are shown for $4$ intervals in stellar mass (rows, as labelled), and as a function of the apparent $r$-band magnitude $m_r$, with fainter galaxies corresponding to lighter-coloured curves. For galaxies with $9<\log_{10}(M_\star/{\rm M}_\odot)<9.5$, there is a bias toward faint, red galaxies (top left panel), while more massive galaxies are biased toward faint, active galaxies (right panels, rows $2$--$4$).}
  \label{FigBias}
\end{figure}

In the context of the quenching analysis, we have checked whether galaxies of a given stellar mass are preferentially included in the sample as a function of either their $(g-r)$ colour or sSFR by comparing how the distributions of these two quantities change as a function of apparent $r$-band magnitude $m_r$, shown in Fig.~\ref{FigBias}. We find that for stellar masses $M_\star>10^{9.5}\,{\rm M}_\odot$, the $(g-r)$ colour distribution is very close to independent of $m_r$, however for lower mass galaxies there is a relative overabundance of faint red galaxies in the catalogue. This is what motivates our stellar mass threshold. Curiously, the sSFR distributions show an opposite behaviour: there is no apperent bias as a function of $m_r$ for low mass galaxies, but there is an overabundance of massive, faint active galaxies. Wishing to use the same galaxy sample for both the colour and sSFR analysis, we could find no way of mitigating both biases simultaneously and so have accepted that the active fractions ($f_{\rm after}$ and $f_{\rm before}$) may be slightly too high overall. We see no reason that, and find no evidence that, these biases should vary with PPS position, so our quenching $t_{\rm mid}$ estimates should be unaffected. We have not identified any biases which depend on the PPS coordinates in the SDSS input catalogue.

The situation for the stripping analysis is somewhat less clear, and more difficult to assess. Whereas in the optical catalogues there are numerous galaxies detected outside our adopted stellar mass and apparent magnitude limits which are useful in assessing possible biases in the catalogue, since we are directly using detection in ALFALA as a tracer of gas content, a similar assessment is more challenging. This is further compounded by the relatively small total number of detections (within the $2.5r_{\rm vir}$ and $2\sigma_{3{\rm D}}$ aperture around our sample of host systems). In a simplified scenario where the presence of a galaxy in the ALFALFA source catalogue depends only on its \HI\ mass and its distance our approach is robust: a galaxy of given stellar mass and in a given group (i.e. at a given distance) has the same probability of being detected regardless of its $R$ and $V$ coordinates, making our measurement of $t_{\rm mid}$ for neutral gas stripping reliable\footnote{We recall that by using detection as a proxy for gas-richness, we must abandon the meaning of $f_{\rm before}$ in the absolute sense: many `gas-rich' galaxies will appear in the catalogue as non-detections simply because they are distant, which will bias $f_{\rm before}$ to lower values.}. In reality, however, other factors influence the detection of sources in the ALFALFA survey. As an example, we consider the possible fates of \HI\ gas which ceases to be detected in a satellite galaxy -- it may have simply been removed, or it may be removed and subsequently ionized, or it may be ionized in place, or it may condense into the molecular phase. Given the poor spatial resolution of the measurements, it is plausible that displaced gas which remains neutral could keep a galaxy above the detection threshold even when the gas is no longer `inside' the satellite. Since the thermodynamic properties of the ambient gas affect the details of how \HI\ disappears from detectability, and vary as a function of PPS and $M_{\rm host}$, our measurements must be affected at some level.

As another example, source confusion in the ALFALFA survey can cause neutral gas in neighbouring galaxies to overlap within the beam ($\sim3.5\,{\rm arcmin}$, or about $90$ -- $250\,{\rm kpc}$ in the redshift interval $z=0.02$ -- $0.06$). This could push a gas-poor galaxy above the detection threshold. This effect is likely more severe for galaxies with low velocity offset $V$ (and also lower radial offset $R$), where satellites are more clustered and confusion with gas associated with the host system itself becomes more likely. Although the overall confusion rate in ALFALFA is $\leq 5$~per~cent \citep{2015MNRAS.449.1856J}, crowding of \HI-bearing galaxies is likely to be more severe in denser regions, especially in the gas-rich group environment. To illustrate this, we consider a $M_{\rm vir}=10^{14}\,{\rm M}_\odot$ host. The $2.5\,r_{\rm vir}$ aperture corresponds to $\sim 3.3\,{\rm Mpc}$, while the $4\sigma_{1{\rm D}}$ velocity aperture\footnote{The aperture is $2\sigma_{1{\rm D}}$ in the absolute value of the velocity difference, but sources will not be confused if their velocity offsets from the host have opposite signs (provided they are well-separated in velocity), so the full $4\sigma_{1{\rm D}}$ applies here.} corresponds to $\sim 1800\,{\rm km}\,{\rm s}^{-1}$. Assuming a fiducial velocity width for satellites of $300\,{\rm km}\,{\rm s}^{-1}$, the PPS aperture around such a host has space for $\sim 1000$ (at $z\sim 0.06$) to $8000$ (at $z\sim 0.02$) uniformly spaced ALFALFA sources without significant confusion between them. Typical hosts of this mass covered by ALFALFA in our sample have $\sim 50$ satellite candidates, of which $\sim 10$ are H\,{\sc i}-detected. A similar calculation for low mass hosts ($M_{\rm vir}=5\times 10^{12}\,{\rm M}_\odot$) gives an estimate of space for about $50$ -- $350$ uniformly distributed satellites, while typical hosts of this mass in our sample have $9$ satellite candidates, of which $2$ are H\,{\sc i}-detected. Given the centrally clustered distribution of satellites, and the additional satellites below our magnitude limit which are not included in our counts, there are likely some confused sources present in our sample despite these estimates, however the majority are likely not confused. Nevertheless, the strong PPS-dependence of this bias motivates some caution \citep[see also][for a complementary assessment of the importance of confusion in dense environments]{2019MNRAS.483.5334S}.

The above discussion of possible biases in ALFALFA serves to highlight some of the trends which should ideally be considered. However, given the limits of the data, we find ourselves unable to fully explore these issues. Our overall assessment is that none of these is likely to drive the several-${\rm Gyr}$ differences in $t_{\rm mid}$ needed to affect our main qualitative conclusions. Nevertheless, we stress that our measurements are a first attempt which can be revisited and improved as larger, more sensitive, and better resolved \HI\ surveys become available.

\subsubsection{Compatibility of orbit libraries}
\label{SubsubsecOLCompat}

One of the key assumptions underpinning our analysis is that the ensemble of orbits drawn from the N-body simulation (see Sec.~\ref{SubsecNbody}) is representative of the ensemble of orbits occupied by the observed galaxies. For instance, satellites heavily stripped of dark matter may still appear as SDSS detections -- the stellar component of galaxies is centrally concentrated and more tightly bound than the bulk of the dark matter -- however, the analogous objects in the N-body simulation (where there is no stellar component) may have their dark matter haloes fully disrupted and thus fail to appear in our list of orbits. This effect is more important for low mass galaxies. In our N-body simulation, haloes of $\log_{10}(M_{\rm vir}/{\rm M}_\odot)<10.5$ are made up of $<50$ particles; below this limit the halo finder begins to struggle to identify them, and of course once only a few particles remain a halo will dissolve, even though a bound core might remain in a realization with higher numerical resolution. A satellite halo which falls in with a mass of $\log_{10}(M_{\rm vir}/{\rm M}_\odot)=11.5$ can therefore be stripped of $\sim 90$~per~cent of its mass before disappearing from the catalogue, while a more massive $\log_{10}(M_{\rm vir}/{\rm M}_\odot)=12.5$ halo would continue to appear until $\sim 99$~per~cent of its initial mass is stripped. We have examined the distribution of stripped mass fractions as a function of maximum halo mass for satellites in our N-body simulation, and find that\footnote{We recall that $M_{\rm sat}$ is defined as the maximum mass which a satellite halo has had at any (past) time.} at $\log_{10}(M_{\rm sat}/{\rm M}_\odot)=12.5$, only $5$~per~cent of (surviving) satellites have been stripped of $>90$~per~cent of their peak mass (the median galaxy with $M_{\rm sat}=10^{12}\,{\rm M}_\odot$ has been stripped of $\sim 35$~per~cent of its mass at $z=0$), even though any stripped of more than this will continue to be tracked while they lose another decade in mass. It is therefore only in the lowest stellar mass bin in our analysis of SDSS galaxies ($M_\star\sim5\times10^9\,{\rm M}_\odot$, corresponding to $M_{\rm sat}\sim3\times10^{11}\,{\rm M}_\odot$ according to our adopted SHMR, see Sec.~\ref{SubsecDataQuenching}) that a significant number ($\sim 20$~per~cent) of orbits will be erroneously missing. These will of course preferentially be the orbits which have the earliest $t_{\rm fp}$'s, which occupy a very biased region of PPS, at low $R$ and low $V$. The net effect on the parameters of our model is to bias $t_{\rm mid}$ to earlier times (lower values). We assess the magnitude of this bias by artificially degrading the resolution of our orbit catalogue by $1\,{\rm dex}$ in mass and repeating our analysis. We find that $t_{\rm mid}$ is underestimated by up to $3\,{\rm Gyr}$ at $M_\star\lesssim5\times 10^{10}\,{\rm M}_\odot$. We stress that this bias only significantly affects the leftmost point on each curve in Fig.~\ref{FigObsFits}, and, encouragingly, in our fiducial measurements these points do not seem to be significantly or systematically offset from those at higher $M_\star$.

We also investigate whether the results presented in Fig.~\ref{FigObsFits} are significantly sensitive to our choice of SHMR. We have repeated our analysis replacing the \citet{2013ApJ...770...57B} SHMR with that of \citet[][one of those most different from that of \citealp{2013ApJ...770...57B} in the recent compilation of \citealp{2019MNRAS.488.3143B}]{2018AstL...44....8K} and find only very small changes in all parameters across all host and satellite masses, e.g. $\lesssim 50\,{\rm Myr}$ difference in the median for $t_{\rm mid}$ and $\lesssim 0.01$ for $f_{\rm before}$.

As a crude upper bound on the systematic error budget due to the compatibility of the orbit libraries with the orbital distribution of observed galaxies, we report the result of our analysis when we match to simulated haloes based on their $z=0$, rather than their peak, mass. This results in essentially all satellites being assigned orbits which should belong to higher mass satellites. Since these preferentially fall in at later times, this results in a large systematic underestimate of $t_{\rm mid}$. However, even this gross mis-assignment of orbits caused an offset of $\lesssim 3\,{\rm Gyr}$ in $t_{\rm mid}$, and occurred nearly uniformly across the entire range in both host and stellar mass, leaving our qualitative conclusions unchanged and lending some additional confidence in their robustness against these types of biases in orbit assignment.

\subsubsection{Statistical considerations}
\label{SubsubsecStatConsid}

Moving on to statistical, rather than systematic, considerations, we performed an additional set of model parameter estimates to check for a `preferred solution' to which the model parameters might converge, for instance due to the mathematical formulation of the model or the choice of priors, rather than being driven by the evidence in the data. We check this by repeating the parameter constraints for the $M_{\rm host}/{\rm M}_\odot>10^{14}$ galaxy sample for the $(g-r)$ colour analysis. However, before evaluating the Markov chains, we randomly `shuffle' the colours of the galaxies within each stellar mass bin, such as to destroy any correlation between galaxy colour and PPS coordinates while preserving all other properties of the galaxy sample. The parameter constraints in this case are characterized by a distribution for $t_{\rm mid}$ which is very broad and prefers large values, specifically extending all the way to the upper limit of the prior distribution ($t_{\rm mid}=10\,{\rm Gyr}$) with either a flat shape at large values, or a peak at the prior bound. This is intuitive: if colour (or any other property) and PPS position are uncorrelated, there is no evidence that the host environment impacts the colour of the satellites -- the very late $t_{\rm mid}$ represents satellites orbiting for a long time within their host without changing their colour. This `preferred solution' seems to be the one reached for the highest stellar mass bin in gas analysis of the $10^{13}<M_{\rm host}/{\rm M}_\odot<10^{14}$ galaxy sample (e.g. rightmost green point in lower left panel of Fig.~\ref{FigObsFits}), and many of the analyses of the $10^{12}<M_{\rm host}/{\rm M}_\odot<10^{13}$ galaxy samples (gas: $4^{\rm th}$, $5^{\rm th}$ and $6^{\rm th}$ stellar mass bins; sSFR and colour: all stellar mass bins). This interpretation is corroborated by the $f_{\rm HI\,detected}$, $f_{\rm active}$ and $f_{\rm blue}$ distributions in PPS for these sets of galaxies (see figures in Appendix~\AppPPSMaps) which do not show a clear gradient in PPS. We therefore do not consider these $t_{\rm mid}$ estimates reliable. In the case of the single analysis in the intermediate host mass sample this does not particularly impact the physical interpretation of our analysis, but for the low host mass sample the implication is that most hosts in this mass range have not yet had time to `fully process' their present-day satellite population -- we will discuss this further in Sec.~\ref{SubsecInterpret} below.

\subsubsection{Realism of the model}

Our model, as summarized in Fig.~\ref{FigSchematic}, clearly cannot be a perfect description of the real time-evolution of the blue/active/gas-rich fraction in dense environments. It does not have enough freedom in shape to accomodate the full spectrum of possibilities: it assumes that the fraction does not evolve outside the time interval defined by $t_{\rm mid}$ and $\Delta t$, and that the times when individual galaxies make the transition from blue/active/gas-rich to red/passive/gas-poor are uniformly distributed over the $\Delta t$ interval, leading to a linear decline. We showed in Sec.~\ref{SubsecModelTests} that we are able to recover all parameters perfectly, within the statistical uncertainties, when the model is an exact description of the data. However, in the more `realistic' test using the Hydrangea clusters, the $\Delta t$ parameter, in particular is not accurately recovered. We have not found any other plausible explanation in the course of the various tests and method variations which we have carried out, so we tentatively attribute this failure to reliably recover the $\Delta t$ timescale to the inevitable mismatch between the form of the model and the underlying `truth' encoded in the data. The mismatch cannot be too severe, however, as evinced by the excellent recovery of the other parameters illustrated in Fig.~\ref{FigModelTest}. We note that this is in part due to a careful formulation of the model -- a mathematically equivalent formulation which replaces $t_{\rm mid}$ and $\Delta t$ with alternative parameters $t_{\rm start}=t_{\rm mid} - \frac{1}{2}\Delta t$ and $t_{\rm end}=t_{\rm mid} + \frac{1}{2}\Delta t$ introduces a strong degeneracy between the two `time' parameters and allows the uncertainty in the width of the transition ($\Delta t$) to wash out the tight constraint on its timing ($t_{\rm mid}$).

\subsubsection{Summary}

Taken together, our assessment of the overall implications of the various biases and uncertainties discussed in this section are:
\begin{itemize}
\item There may be significant systematic offsets in $t_{\rm mid}$, of up to $\sim 3\,{\rm Gyr}$, but we find that most such possible offsets tend to apply approximately uniformly at all stellar masses and host halo masses. This implies that our recovery of the ordering of the transitions -- gas-rich to gas-poor, followed significantly later by active to passive, and then almost immediately by blue to red -- is most likely a robust result. We are similarly confident in our conclusion that gas stripping and quenching proceed somewhat more quickly in more massive host haloes.
\item The highest stellar mass bin in the gas stripping analysis for the intermediate host mass bin appears to correspond to the `preferred solution' of the model in the absence of evidence and are unlikely to represent reliable measurements of the model parameters. This is corroborated by the absence of a visible gradient in $f_{{\rm HI}\,{\rm detected}}$ in PPS. The same `preferred solution' is also found for most constraints of our model parameters for the low host mass sample (all tracers); we return to this point in Sec.~\ref{SubsecInterpret}.
\item The limited resolution of the N-body simulation, which causes low-mass satellite haloes to be disrupted too early, is likely to be driving an underestimate of $t_{\rm mid}$ in the lowest stellar mass bin of each host halo mass bin. This could mask a decreasing trend of $t_{\rm mid}$ with increasing stellar mass, a point to which we will return in Sec.~\ref{SubsecCompare}, below.
\end{itemize}

\subsection{Comparison with prior studies}
\label{SubsecCompare}

\subsubsection{\citet[][\citetalias{2016MNRAS.463.3083O}]{2016MNRAS.463.3083O}}
\label{SubsubsectionCompareOH16}

In terms of methodology, the previous study most similar to ours is that of \citetalias{2016MNRAS.463.3083O}. The first crucial difference between the two analyses is that they use the infall time (defined as the first inward crossing of $2.5r_{\rm vir}$) as a reference `$t=0$' from which to measure the timescale for quenching, while we use the time of the first pericentric passage. We would therefore expect our $t_{\rm mid}$ values, specifically for the sSFR input, to be smaller than theirs by approximately the time taken for a satellite to orbit from infall to first pericentre, about $3$--$4\,{\rm Gyr}$. However, comparing their measurements of $t_{1/2}$, also the time when half of the satellite population has `transitioned', to our $t_{\rm mid}$ (the dashed and solid red lines in their fig.~9 correspond closely in terms of galaxy selection to the green and red dashed lines in our Fig.~\ref{FigObsFits}), our measurements are at most $\sim 1\,{\rm Gyr}$ shorter. In order to unambiguously determine the origin of this apparent discrepancy, we have repeated our analysis with four changes, implemented one at a time, which together lead to a quantitative reproduction of the result of \citetalias{2016MNRAS.463.3083O}. The first two, which change our measurement very little, are to adopt their stellar mass binning, and replace our linear decline model (Eq.~\ref{EqLinearDrop}) with their exponential decline model (their eq.~12). Next, we replace the probability distributions used to estimate $t_{\rm fp}$ with similar distributions to estimate the infall times of satellites. This results in the expected $\sim 3.5\,{\rm Gyr}$ upward shift in $t_{\rm mid}$. Finally, we reproduce their use of the $z=0$ satellite halo masses in the N-body simulation to associate possible orbits to observed satellites (whereas we use the maximum past halo mass of satellite haloes), which results in a $\sim 2.5\,{\rm Gyr}$ downward shift in $t_{\rm mid}$. These changes together bring the two analyses into agreement at $\log{10}(M_\star/{\rm M}_\odot)\gtrsim 10.5$; the remaining discrepancy at the low mass end is consistent with being due to their (erroneous) use of a lower resolution N-body simulation (see Sec.~\ref{SubsubsecOLCompat}). In summary, our finding that the characteristic time for quenching, $t_{\rm mid}$, is several ${\rm Gyr}$ after the first pericentric passage, rather than around or just after this even as reported by \citetalias{2016MNRAS.463.3083O}, is due to improvements in model assumptions, and that the present study should be taken as superseding this earlier measurement.

\subsubsection{\citet[][\citetalias{2013MNRAS.432..336W}]{2013MNRAS.432..336W}}
\label{SubsubsecCompareW13}

We next compare our measurement to the conclusions of \citetalias{2013MNRAS.432..336W}, who presented the first empirical evidence for the `delayed-then-rapid' quenching scenario. We focus in particular on comparison with this study (i) because the differences we explain below serve to highlight many important systematic dependencies on model assumptions and (ii) because it is recognized as a landmark result of the field. Our analysis uses the same underlying optical survey (SDSS DR7 spectroscopic sample) as theirs, although with a different group catalogue \citep[they use that of][]{2012MNRAS.424..232W}. Our methodology differs from theirs on a few key points. First, whereas the `rapid' nature of the blue-to-red/active-to-passive/gas-rich-to-gas-poor transition is as assumption\footnote{Actually, we only assume that galaxies cross our sharp deliniation of blue/red (or similar) rapidly, so we are not sensitive to the equivalent of the `fading time' of \citetalias{2013MNRAS.432..336W}, which is constrained mostly by the shape of the sSFR distribution, which we have reduced to the fraction $f_{\rm active}$.} in our method, \citetalias{2013MNRAS.432..336W} constrain the `fading timescale' ($\tau_{\rm Q,fade}$ in their notation), on which the sSFRs of individual satellites decline, directly. $\tau_{\rm Q,fade}$ should not be confused with our $\Delta t$, which represents the interval over which a collection of satellites each cross our sharp division of the population. Second, in their analysis, `infall' is defined as the time when a galaxy first becomes a satellite (defined via friends-of-friends association with a more massive system) of \emph{any} host, in contrast to our use of the first pericentric passage within the (progenitor of) the \emph{current} host system. Our initial expectation is then that our measurements of $t_{\rm mid}$ (for our sSFR analysis), loosely comparable to their $t_{\rm Q}$ (not $t_{\rm Q,delay}$), should be uniformly shorter, as their delay time includes the time in any previous hosts, and the time to orbit from infall to first pericentre.

The fact that our $t_{\rm mid}$ measurements (Fig.~\ref{FigObsFits}, bottom centre) are similar to or larger than their $t_{\rm Q}$ measurements (their fig.~8, upper panel) therefore merits careful consideration. First, we note that the host-mass ordering of the curves in the two figures differs: whereas we find a monotonic increase in $t_{\rm mid}$ with decreasing $M_{\rm host}$ at low $M_\star$, \citetalias{2013MNRAS.432..336W} find the longest $t_{\rm Q}$ in intermediate mass hosts; both figures agree that the delay time is approximately constant with $M_{\rm host}$ at high $M_\star$. The reason for this difference is clear from their fig.~2: in higher mass hosts, the first infall is much earlier than the most recent infall, while in low mass hosts this difference is very small. An approximate correction for this would involve shifting their low/intermediate/high host mass measurements down by $\sim 0.5$/$2$/$3\,{\rm Gyr}$, respectively, which results in the same host-mass ordering as for our measurement. We therefore agree that the \emph{isolated} environmental influence (i.e. excluding group pre-processing) of more massive hosts is felt more quickly by their satellites \citepalias[for further discussion see also][sec.~4.3.1]{2013MNRAS.432..336W}.

The next clear difference is the slope of the curves -- whereas we find generally near-flat dependences of $t_{\rm mid}$ on $M_\star$, \citetalias{2013MNRAS.432..336W} find a strongly declining slope. The most plausible explanation for this apparent discrepancy lies in the differences between our respective treatments of group pre-processing. Whereas \citetalias{2013MNRAS.432..336W} `start the clock' for the quenching delay time when a galaxy first becomes a satellite of any host, we use the first pericentric passage in the present-day host as a reference time. For massive satellite galaxies, these two definitions turn out to be at least roughly equivalent (except for the time interval required to orbit from infall to pericentre): the majority of more massive satellites fall into their hosts at later times (by about $1.5\,{\rm Gyr}$ between $\log_{10}(M_\star/{\rm M}_\odot)=9.7$ and $11.3$) and, where they are quenched as satellites, are quenched in their present-day host \citepalias[][especially their fig.~10]{2013MNRAS.432..336W}. Lower mass galaxies, however, tend to have earlier `first infall' times (e.g. into a group that will later fall into a cluster), such that even if they reach their final host un-quenched, they are already partially `processed' and more vulnerable to the environmental influence of the present-day host. This `pre-processed' population will have preferentially shorter quenching times (as defined in terms of time spent in the present-day host), flattening out the trend with $M_\star$ relative to the measurement of \citetalias{2013MNRAS.432..336W}.

Differences in the trend with $M_{\rm host}$ and $M_\star$ now having been explained, we are left with any differences in the absolute normalization of $t_{\rm mid}$ \citepalias[or, in the notation of][$t_{\rm Q}$]{2013MNRAS.432..336W} left to be accounted for. The easiest sub-sample of galaxies to use for this is the high-$M_\star$ end of the satellites with $M_{\rm host}/{\rm M}_\odot>10^{14}$. Here, offsets due to the different definitions of infall time, and different handling of group pre-processing, should be minimal, as outlined above. We would then expect our measurement to be perhaps $2\,{\rm Gyr}$ shorter than that of \citetalias{2013MNRAS.432..336W}, to account for the time to orbit from infall (defined in terms of a friend-of-friends membership criterion) to first pericentre. However, the difference runs in the opposite sense, with our typical quenching time lagging theirs by $\sim 2\,{\rm Gyr}$ (for this particular $M_{\rm host}$ and $M_\star$). We have found no entirely satisfactory explanation for this discrepancy, and will return to this point in the summary (Sec.~\ref{SubsubsecComparisonSummary}) below.

\subsubsection{\citet{2020ApJS..247...45R}}

\citet{2020ApJS..247...45R} use the SFR distribution of disc galaxies as a function of their PPS coordinates to derive a quenching timescale, under the assumption that the earliest infalling galaxies have the lowest star formation rates. As in our analysis, they consider the environmental influence of the present-day host independent of (or, perhaps more accurately, over-and-above) pre-processing. However, rather than an explicit definition of quenching, they adopt a simple model for the star formation history as a function of time since infall in order to constrain two exponential decay timescales for the SFR, corresponding to outside ($\tau_{\rm ex-situ}$) and inside ($\tau_{\rm cluster}$) the final host, and a delay timescale ($t_{\rm d}$) representing the time since infall before the transition between the two decay timescales occurs\footnote{They also constrain a parameter $\alpha$ describing the redshift evolution of the cluster quenching timescale, but this turns out to be consistent with no evolution in all cases (although the confidence intervals are rather broad).}. \citet{2020ApJS..247...45R} helpfully provide their measurements following the definitions of \citetalias{2016MNRAS.463.3083O} in their fig.~11; the $t_{\rm Q}$ given there is directly comparable to our $t_{\rm mid}$, except for an offset to higher $t_{\rm Q}$ corresponding to the typical difference between the infall (defined at $2.5r_{\rm vir}$) and first pericentre times, which is $3.8\,{\rm Gyr}$ (independent of $M_\star)$, for a host mass range $5\times 10^{13}<M_{\rm host}/{\rm M}_\odot<10^{15}$ -- most similar to the red line in the bottom centre panel of Fig.~\ref{FigObsFits}. It is immediately clear that our $t_{\rm mid}$ well exceeds their (adjusted) $t_{\rm Q}$ at all stellar masses -- by $\sim 2\,{\rm Gyr}$ at the low stellar mass end, up to $\sim 5\,{\rm Gyr}$ at the high stellar mass end. This is further exacerbated by the fact that our $t_{\rm mid}$ likely corresponds to that for galaxies just now quenching\footnote{Suppose that active galaxies in our sample which are just now falling into groups will become passive (arbitrarily) $1\,{\rm Gyr}$ earlier than would be expected from our measured $t_{\rm mid}$. Since the delay time has not yet elapsed for these objects, there is no evidence for this change in timescale contained in the data -- our measurement cannot be sensitive to it.}, whereas \citet{2020ApJS..247...45R} estimate the $z\sim0$ value of $t_{\rm Q}$ assuming a $t_{\rm Q}(z_{\rm inf})\propto(1+z_{\rm inf})^{-1.5}$ dependence on the infall redshift\footnote{This is very close to $t_{\rm Q}(t_{\rm inf})\propto t_{\rm inf}$, where $t$ is the age of the Universe.}, motivated by the redshift dependence of the crossing timescale of host systems. We might therefore expect our quenching timescale to be about half of their measurement, though this is difficult to ascertain precisely given the substantial ambiguity in defining the reference time from which any delay should be measured.

\subsubsection{Summary}
\label{SubsubsecComparisonSummary}

At a glance, the apparent quantitative agreement -- once differences due to the various definitions adopted are accounted for -- between the quenching timescales measured by \citetalias{2013MNRAS.432..336W}, \citetalias{2016MNRAS.463.3083O} and \citet{2020ApJS..247...45R} would suggest that our measurements are outliers and perhaps suspect \citep[though there are also some measurements of longer quenching timescales potentially consistent with ours, e.g.][]{2014MNRAS.440.1934T,2014MNRAS.442.1396W}. However, as explained in Sec.~\ref{SubsubsectionCompareOH16}, by systematically adjusting one aspect of our analysis at a time until we adopt an identical set of assumptions to \citetalias{2016MNRAS.463.3083O}, we can reproduce their measurements in quantitative detail. This approach reveals that the discrepancy can be explained in terms of (i) their assignment of orbits to satellites neglecting tidal stripping of dark matter, and (ii) their (erroneous) use of an N-body simulation with inadequate numerical resolution, suggesting that our measurements are not `simply outliers'. We are further encouraged by the excellent agreement between the `true' $t_{\rm mid}$ values -- which are fully independent of the $t_{\rm fp}$ probability distributions derived from the N-body simulation -- and the estimates derived from `observed' quantities, as shown in Fig.~\ref{FigModelTest}.

Nevertheless, making the link between the stellar mass of satellites and which candidate haloes' orbits should be selected based on their mass in the N-body simulation remains the most challenging aspect of our approach, and directly impacts the absolute calibration of $t_{\rm mid}$. \citetalias{2013MNRAS.432..336W} and \citet{2020ApJS..247...45R} also cite difficulties in this area. \citetalias{2013MNRAS.432..336W} discuss at length (e.g. their appendix~A) the assumptions required to account for the continued stellar mass growth of satellites while their dark matter haloes are being tidally stripped, and \citet{2020ApJS..247...45R} need to work around a factor of $\sim2$ mismatch in the satellite stellar mass function between their sample of observed satellites and their cosmological hydrodynamical simulations of clusters. A fully self-consistent treatment requires simultaneously accounting for: (i) the dark matter mass loss after infall; (ii) stellar mass growth due to continued star formation between infall and quenching; (iii) stellar mass loss due to tides \citep[though this is likely to be a minor effect, see][]{2017MNRAS.464..508B} and supernovae/winds from massive stars. In addition to varying strongly as a function of the (unknown) stellar (or halo) mass at infall, and the (unknown) infall time via the redshift-dependence of the stellar-to-halo mass relation, the stellar mass growth in particular also depends on the quenching timescale. These strong couplings between unknown parameters require either additional assumptions, or an increase in the dimensionality of the parameter space through the introduction of additional `nuisance parameters'. Realistic implementations of both approaches in our framework would be associated in a significant increase in computational cost for each model evaluation, motivating us to proceed with our simplified approach, for the present.

With the above in mind, we focus our interpretation below on what we perceive as the strengths of our analysis: a uniform input galaxy sample across multiple tracers (colour, Balmer emission lines and \HI) yielding robust estimates of the relative timescales associated to quenching and stripping as traced by each, and the use of the first pericentre as a reference time offering better discrimination between processes which occur at or well away from first pericentre than an `infall' reference time.

\subsection{Physical interpretation}
\label{SubsecInterpret}

To guide our interpretation, we begin with an illusturative example. According to our parameter constraints, a galaxy of $M_\star\sim 10^{10}\,{\rm M}_\odot$ in a massive host system ($M_{\rm host}>10^{14}\,{\rm M}_\odot$) typically continues star formation for $\sim 3\,{\rm Gyr}$ after it is no longer detected in ALFALFA. To remain above our threshold sSFR of $10^{-11}\,{\rm yr}^{-1}$ over this interval, it must grow its stellar mass by at least $\Delta M_\star\sim 3\times 10^8\,{\rm M}_\odot$. While a typical galaxy with this stellar mass that is detected in ALFALFA has $M_{\rm HI}\sim 10^{10}\,{\rm M}_\odot$, by the time it is undetected it likely has $M_{\rm HI}\lesssim 5\times 10^9\,{\rm M}_\odot$, so the star formation `efficiency' during this time interval must be $\Delta M_\star/M_{\rm HI}\gtrsim 0.1$. At a glance, this value suggests that the gas supply is ample to fuel the required star formation. However, the sSFR is unlikely to remain just at the threshold value -- more likely the average over the $3\,{\rm Gyr}$ interval is a factor of a few higher. Furthermore, star formation is not a highly efficient process: as stars are formed, some gas is launched as a wind with a mass loading $\eta=\dot{M}_{\rm wind}/\dot{M}_\star$. Within a cluster, the gas in the wind is unlikely to be able to return to fuel later star formation as any wind launched beyond the ISM is exposed to ram pressure from the harsh intra-cluster medium. The value of $\eta$ is debated, but is likely $\sim 1$ for galaxies of this mass \citep[e.g.][and references therein]{2014MNRAS.442L.105M}, so that at most $\sim \frac{1}{2}$ the available gas mass can ultimately end up in stars. Together, these considerations leave rather little room for H\,{\sc i} gas to be removed by other processes, such as ram pressure or tides, over the same time interval.

It is clear that the ISM gas feels the influence of such massive host systems relatively early: there are numerous examples of satellites with prominent gas tails generally agreed to be on their first approach to their host \citep[e.g.][]{2009AJ....138.1741C,2010MNRAS.408.1417S,2017ApJ...838...81Y,2018MNRAS.476.4753J,2020MNRAS.495..554R}. One physical picture consistent with this, and our measurements, is that as a satellite falls into a cluster, it is stripped of H\,{\sc i} by ram pressure on its initial approach -- not completely, but enough that it is no longer detected in ALFALFA by the time it reaches pericentre. RPS then ceases as soon as the satellite passes pericentre and the ram-pressure force rapidly drops off. The remaining, centrally concentrated H\,{\sc i} then continues to fuel star formation, driving winds which carry away some of the gas. It is unclear whether it is the eventual depletion of the gas supply by this process, or a later episode of stripping, which finally quenches the galaxy. Indeed, based on the timescales involved and the scatter in galaxy and orbital properties, both of these possibilities probably occur with non-negligible frequency. This picture is also consistent with the observations discussed in \citet{2019ApJ...873...52O}, who found a significant population of cluster satellites, likely on their first orbital passage, with recently quenched outer discs but continued star formation in their central regions.

We have so far neglected another reservoir of fuel for star formation to which the observations we have used in our analysis are blind: the molecular gas phase. Around the time of infall, this reservoir is unlikely to contain enough additional gas to make much difference to the picture outlined above -- the molecular-to-atomic gas ratio of our example galaxy would be of about $0.15$ \citep{2018MNRAS.476..875C}. However, \citet{2020ApJ...897L..30M} suggest that atomic hydrogen may be highly efficiently converted to ${\rm H}_2$ by ram pressure-driven compression in `jellyfish' galaxies. This leaves the detailed evolution of the satellite galaxies somewhat ambiguous -- if a substantial amount of H\,{\sc i} is fed into the molecular reservoir in this way, then the long delay before quenching becomes less surprising, and subsequent RPS or tidal stripping may even be \emph{required} to truncate star formation. The ambiguity could be alleviated by (i) spatially resolved observations of the H\,{\sc i} gas sufficiently deep to reveal the extent to which RPS removes atomic gas, e.g. by measuring how much is present in a stripped tail of gas \citep[see][for one example where the majority of H\,{\sc i} is contained in a tail]{2020MNRAS.494.5029D}, and (ii) observations of molecular gas tracers, such as dust or CO \citep[e.g. the forthcoming VERTICO survey of CO in Virgo cluster satellites, see][]{2020sea..confE..13B}, to determine when along satellite orbits the content of this reservoir increases or is depleted.

Turning our attention next to lower mass hosts, we consider the same example satellite galaxy around a $10^{13}$ to $10^{14}\,{\rm M}_\odot$ host. Such satellites quench slightly ($\sim 2\,{\rm Gyr}$) later than those around more massive hosts, but become undetected in ALFALFA much later ($\sim 3$ to $4\,{\rm Gyr}$, and thus \emph{after} the first pericentric passage\footnote{Though this point is difficult to establish conclusively, as it is sensitive to the treatment of $f_{\rm after}$, see Appendix~\AppExtRes.}), such that the interval between these two events is shorter, about $2\,{\rm Gyr}$. Together, this suggests that RPS is insufficient to push these galaxies below the ALFALFA detection threshold on their first passage through their host. That the transition from gas-rich to gas-poor occurs near the first apocentric passage, when the RPS rate is likely to be near zero -- any gas that could be stripped should already have been removed when the ram pressure peaked during the first pericentric passage -- suggests that it is more likely consumption of gas by star formation, rather than removal through stripping, which carries these satellites across the ALFALFA detectability threshold. Conversely, that these galaxies quench only $\sim 2\,{\rm Gyr}$ later leaves perhaps too little time to consume \emph{all} the remaining gas in forming stars, suggesting that the second peak in ram pressure and tidal forces at the second pericentric passage may be the ultimate driver of the shutdown of star formation in these satellites.

The situation in the lowest mass hosts is more ambiguous. The statistical errors are larger, and $t_{\rm mid}$ becomes very long ($\sim 8\,{\rm Gyr}$ for the transition in colour, with confidence intervals pushing against the upper bound of our adopted prior of $0<t_{\rm mid}/{\rm Gyr}<10$), i.e. an appreciable fraction of the age of the Universe. We recall that our methodology is only sensitive to transitions which have actually occured by the time of observation -- essentially, it finds the time along the orbits of satellites \emph{actually present in their hosts} when there is a transition in the gas content/SFR/colour of the satellites. If this transition has yet to occur, our choice to fix $f_{\rm after}=0$ is incorrect. In the case of $10^{12}$ to $10^{13}\,{\rm M}_\odot$ hosts, when $f_{\rm after}$ is left free to vary (see Appendix~\AppExtRes), $f_{\rm after}$ shows some preference for non-zero values\footnote{In this case, the $t_{\rm mid}$ values drop by $1$ to $2\,{\rm Gyr}$ across all tracers.}, suggesting that the satellite population has not yet had time to be completely `processed' by these hosts. Put another way, a similar fraction of satellites with PPS coordinates consistent with very early first pericentric passages, e.g. $\gtrsim 8\,{\rm Gyr}$ ago, remain gas rich and star forming as to what is found in the field, suggesting that these low-mass groups are extremely inefficient at stripping and quenching their satellites. This interpretation is also consistent with the absence of clear gradients in $f_{\rm blue}$, $f_{\rm active}$ and $f_{\rm HI\,detected}$ as a function of the PPS coordinates for satellites in these low-mass hosts (see Appendix~\AppPPSMaps). We note that lower mass groups are \emph{more} effective at destroying satellites through mergers with the centrals, so many satellites may not actually survive long enough to be stripped and quenched by the intra-group medium. Merged satellites do not appear in our observational sample, nor do they contribute to our orbital parameter probability distributions, so this effect is not captured in our analysis.

Finally, we comment briefly on the trends in our $t_{\rm mid}$ measurements as a function of $M_\star$. For colour and sSFR, the trend is rather flat (or slightly decreasing with increasing $M_\star$), across all host masses. We recall that this is intimately connected to our treatment of `pre-processing' (see Sec.~\ref{SubsubsecCompareW13}). This flatness leads us to two conclusions. First, it suggests that, although lower-mass hosts can eventually quench their satellites, if these fall into a more massive host, the influence of the new, denser environment will usually be sufficiently harsher that it will itself set the timescale for quenching, regardless of the earlier processing. Second, it seems that it is ultimately the external influence of the host system which determines these timescales, independent of the mass of the satellite, unless it arises as a result of some fine-tuning between the scalings of sSFR, gas fraction, and other galaxy properties with stellar mass.

The trend in $t_{\rm mid}$ for gas, on the other hand, appears to be slightly increasing with increasing $M_\star$. This seems consistent with a scenario where lower $M_\star$ galaxies are stripped of gas (by ram pressure or tides). In the absence of such stripping, galaxies at the low-mass end of our sample range typically have enough gas to sustain star formation for a Hubble time or more, while those at the high-mass end have lower gas consumption timescales. In the absence of significant stripping, we would therefore expect a decreasing trend in $t_{\rm mid}$ with increasing $M_\star$. The increasing trend instead evokes a picture where the deeper potential wells of more massive satellites allow them to retain their gas somewhat longer as the relative impact of ram pressure and tides are weaker.

\section{Summary and Conclusions}
\label{SecConc}

We have used a combination of galaxy photometry and spectroscopy from the SDSS, \HI\ fluxes from the ALFALFA survey, and orbital information inferred from tracking haloes in an N-body simulation to constrain a simple model linking the star formation and gas properties of satellites with their orbital histories. In the context of discriminating between the various physical processes contributing to the shutdown of star formation in satellites, the $t_{\rm mid}$ and $\Delta t$ parameters of our model are the most informative. $t_{\rm mid}$ is the median time relative to the first pericentric passage within the present-day host when satellites transition from blue to red, from active to passive, or from gas-rich (defined as detected in the ALFALFA survey) to gas-poor, while $\Delta t$ describes the (full width) scatter in this transition time; this latter parameter is poorly constrained in our analysis and treated as a `nuisance parameter'. Built into our methodology through our fiducial choice of $f_{\rm after}=0$ is the assumption that, provided it orbits within its host for `long enough' (as expressed by the parameter combination $t_{\rm mid}+\frac{1}{2}\Delta t$), all satellites are ultimately stripped of gas and are quenched. Our main results are summarized as follows:
\begin{itemize}
\item We clearly detect the separation in time between the three stages of environmental processing probed by our analysis: first, neutral gas ceases to be detected, followed a few ${\rm Gyr}$ later by a drop in sSFR (traced by the disappearance of Balmer emission lines associated to star formation), and in turn $\lesssim 1\,{\rm Gyr}$ later by reddening in $(g-r)$. This ordering is ubiquitous across hosts from $M_{\rm host}=10^{12}$ to $\sim10^{15}\,{\rm M}_\odot$, though the time intervals between the stages are somewhat longer in more massive hosts.
\item At fixed host mass, $t_{\rm mid}$ associated to each tracer (\HI, sSFR, colour) is remarkably independent of satellite stellar mass. We note that this is intimately connected to our treatment of `pre-processing'. If we instead measured $t_{\rm mid}$ relative to the first pericentric passage in \emph{any} host, rather than the \emph{current} host \citep[i.e. similar to][though they use infall time, not pericentre time]{2013MNRAS.432..336W}, we would expect to find overall higher values for $t_{\rm mid}$. Furthermore, low-mass galaxies would be offset by more than high-mass galaxies (see Sec.~\ref{SubsubsecCompareW13}), which would result in a (stronger) negative gradient in $t_{\rm mid}$ as a function of $M_\star$.
\item In massive hosts ($M_{\rm host}\gtrsim 10^{14}\,{\rm M}_\odot$), neutral hydrogen disappears before or around the first pericentric passage, while in lower mass hosts it remains detectable up to several ${\rm Gyr}$ later. This difference is not driven by a distance or similar bias in the sample of host systems.
\item Star formation persists in a typical satellite for up to $\sim5\,{\rm Gyr}$ after the first pericentric passage, by which time most satellites will be somewhere between second pericentre and having completed multiple orbits within the host.
\item In low-mass hosts ($10^{12}<M_{\rm host}/{\rm M}_\odot<10^{13}$), we find very large values of $t_{\rm mid}$. Coupled with a lack of a clear gradient in the blue, active, and \HI-detected fractions as a function of PPS position, and a preference for $f_{\rm after}>0$ when our fiducial choice of $f_{\rm after}=0$ is relaxed, this suggests that such low-mass groups have been very inefficient at stripping and quenching their present-day satellite population (see Secs.~\ref{SubsubsecStatConsid} and \ref{SubsecInterpret}).
\end{itemize}

Our measurements are broadly consistent with the `delayed-then-rapid' quenching scenario \citep{2012MNRAS.423.1277D}, however we infer a delay timescale which is much longer than most other studies \citep[\citetalias{2013MNRAS.432..336W}; \citetalias{2016MNRAS.463.3083O};][in particular]{2020ApJS..247...45R}. Detailed tests of our methodology on a mock galaxy sample drawn from the Hydrangea cosmological hydrodynamical simulation suite suggest that our recovery of $t_{\rm mid}$ is accurate (see Sec.~\ref{SubsecModelTests}), but the absolute calibration of this timescale, and comparison across studies employing various zero-points from which the delay is measured, remains challenging.

Our measurement of a long quenching timescale contrasts with results from contemporary galaxy formation simulations, which predominantly find that hosts -- especially rich clusters -- quench their satellites rather efficiently on the first orbital passage (e.g. \citealp{2017MNRAS.470.4186B} with the Hydrangea suite -- see also Fig.~\ref{FigParamSelection}, \citealp{2019MNRAS.488.5370L} with the Magneticum suite, \citealp{2019MNRAS.484.3968A} with the `TheThreeHundred' suite), but the leading reason for this is likely numerical: the cold ISM is notoriously difficult to capture accurately in such simulations \citep{2017MNRAS.470.4186B}. Alternately, the simulated galaxies may have lower gas content than their observed counterparts, making them more susceptible to earlier quenching \citep[][but see also \citealp{2016MNRAS.456.1115B} and \citealp{2019MNRAS.487.1529D} who argue that EAGLE and IllustrisTNG galaxies have gas fractions in broad agreement with observed values]{2015MNRAS.447L...6S}.

It may be possible to further improve the general approach of constraining models for stripping and quenching via inference of the orbits of satellites based on simulation `orbit libraries'. For instance, the distribution of satellite orbits around host systems in different dynamical states differ \citep{2020MNRAS.492.6074H}; this could be incorporated into the model constraints based on observable properties of the host systems. The PPS coordinates of satellites also encode additional information on their orbits, for instance they are (weakly) sensitive to the minimum separation from the host; our model could be extended to incorporate this extra information. Alternatively, or in addition, a more sophisticated model describing the redshift evolution of the satellite and/or host properties could be used, and additional constraints such as the full colour or sSFR distributions could replace our simple, binary classifications -- this would amount to an attempt to combine the strengths of several existing studies. However, our impression is that the required sophistication of the models begins to be cumbersome, and that it may be more fruitful to adopt a more direct simulation approach. While the detailed structure of the multiphase ISM and its hydrodynamical interactions with the intra-group/cluster medium remain out of reach of the resolving power of current cosmological hydrodynamical simulations, a semi-analytic or hybrid (e.g. a semi-analytic model implemented on top of a hydrodynamical rather than an N-body simulation) approach may be a more straightforward framework in which to attempt to capture the full breadth of environmental physics affecting the evolution of satellite galaxies in a fully self-consistent manner.

\section*{Acknowledgements}\label{sec-acknowledgements}

We thank the anonymous referee for a very detailed and constructive review.

Thank you to S. Ellison for assistance with the SDSS stellar masses and SFRs.

KAO and MAWV acknowledge support by the Netherlands Foundation for Scientific Research (NWO) through VICI grant 016.130.338 to MAWV. KAO acknowledges support by the European Research Council (ERC) through Advanced Investigator grant to C.S. Frenk, DMIDAS (GA 786910). YMB acknowledges funding from the EU Horizon 2020 research and innovation programme under Marie Sk\l{}odowska-Curie grant agreement 747645 (ClusterGal) and the Netherlands Organisation for Scientific Research (NWO) through VENI grant 639.041.751. JH acknowledges research funding from the South African Radio Astronomy Observatory, which is a facility of the National Research Foundation, an agency of the Department of Science and Innovation. MJH acknowledges an NSERC Discovery Grant. This research has made use of NASA's Astrophysics Data System.

This research was supported by the Munich Institute for Astro- and Particle Physics (MIAPP) which is funded by the Deutsche Forschungsgemeinschaft (DFG, German Research Foundation) under Germany's Excellence Strategy - EXC-2094 - 390783311.

The Hydrangea simulations were in part performed on the German federal maximum performance computer “HazelHen” at the maximum performance computing centre Stuttgart (HLRS), under project GCS-HYDA / ID 44067 financed through the large-scale project “Hydrangea” of the Gauss Center for Supercomputing. Further simulations were performed at the Max Planck Computing and Data Facility in Garching, Germany.

Funding for the Sloan Digital Sky Survey (SDSS) has been provided by the Alfred P. Sloan Foundation, the Participating Institutions, the National Aeronautics and Space Administration, the National Science Foundation, the U.S. Department of Energy, the Japanese Monbukagakusho, and the Max Planck Society. The SDSS Web site is \url{www.sdss.org}.

The SDSS is managed by the Astrophysical Research Consortium (ARC) for the Participating Institutions. The Participating Institutions are The University of Chicago, Fermilab, the Institute for Advanced Study, the Japan Participation Group, The Johns Hopkins University, Los Alamos National Laboratory, the Max-Planck-Institute for Astronomy (MPIA), the Max-Planck-Institute for Astrophysics (MPA), New Mexico State University, University of Pittsburgh, Princeton University, the United States Naval Observatory, and the University of Washington.

\section*{Data availability}

The SDSS Data Release 7 is available at \url{skyserver.sdss.org/dr7}. The added value catalogues with stellar masses are available from the VizieR service (\url{vizier.u-strasbg.fr}), catalogue entry \verb!J/ApJS/210/3!. The star formation rates are available from \url{wwwmpa.mpa-garching.mpg.de/SDSS/}. The two group catalogues are available via VizieR (entry \verb!J/MNRAS/379/867!) and \url{gax.sjtu.edu.cn/data/Group.html}, respectively. The $\alpha.100$ data release of the ALFALFA survey, including the optical counterparts from \citet{2020arXiv201102588D}, is available at \url{egg.astro.cornell.edu/alfalfa/data/}. An imminent public release of the Hydrangea simulation data is being prepared, contact YMB (\href{mailto:bahe@strw.leidenuniv.nl}{bahe@strw.leidenuniv.nl}) for details. The VVV simulation initial conditions and snapshots are not currently publicly available. Enquire with the VVV authors, or any N-body simulation with a similar cosmology and resolution could be substituted and expected to give statistically equivalent results. Satellite orbit and interloper data tables based on the VVV simulation are available on request from KAO (\href{mailto:kyle.a.oman@durham.ac.uk}{kyle.a.oman@durham.ac.uk}), or may be created for any N-body simulation using the publicly available {\sc rockstar} (\url{bitbucket.org/gfcstanford/rockstar}), {\sc consistent-trees} (\url{bitbucket.org/pbehroozi/consistent-trees}) and {\sc orbitpdf} (\url{github.com/kyleaoman/orbitpdf}) codes. An implementation of the statistical models is available on request from KAO. The marginalized 68~per~cent confidence intervals for the model parameters in Fig.~\ref{FigObsFits} and Figs.~\AppExtRes{}1--\AppExtRes{}4 are tabulated in Appendix~\AppTables; full Markov chains are available on request from KAO.

\bibliography{paper}

\appendix

\section*{Appendices}
\label{SecApp}
We include as supplementary materials, available from the publisher:
\begin{itemize}
\item Appendix~\AppPPSMaps: Figures similar to Fig.~\ref{FigDataIn} for all bins in host mass and satellite stellar mass.
\item Appendix~\AppExtRes: Expanded version of Fig.~\ref{FigObsFits} including $\Delta t$, and alternative version of Fig.~\ref{FigObsFits} where $f_{\rm after}$ is a free parameter.
\item Appendix~\AppAltGas: Re-derivations of our main results assuming different definitions of what constitutes a `gas rich' galaxy. 
\item Appendix~\AppExtSim: Figures similar to Figs.~\ref{FigFitIllustrate} and \ref{FigCornerExample} for the other stellar mass bins included in Fig.~\ref{FigModelTest}.
\item Appendix~\AppTables: Tables with the median and 68~per~cent confidence intervals for all the marginalized posterior probability distributions shown in the supplementary Figs.~\AppExtRes{}1--\AppExtRes{}4 (i.e. including those from Fig.~\ref{FigObsFits}).
\item Appendix~\AppCorners: Figures showing the individual and pairwise marginalized probability distributions (i.e. similar to Fig.~\ref{FigCornerExample}) for the parameters summarized in Fig.~\ref{FigObsFits}, and similar constraints when $f_{\rm after}$ is a free parameter.
\end{itemize}

\label{lastpage}

\end{document}